\documentclass[a4paper,11pt]{article}

\usepackage{jheppub}

\title{\boldmath Radiated momentum in the Post-Minkowskian \\ worldline approach via reverse unitarity}

\author{Massimiliano Maria Riva}
\author{and Filippo Vernizzi}
 \affiliation{Institut de physique th\' eorique, Universit\'e  Paris Saclay, CEA, CNRS, \\ 91191 Gif-sur-Yvette, France}
 
\emailAdd{massimiliano-maria.riva@ipht.fr}
\emailAdd{filippo.vernizzi@ipht.fr }

\usepackage{adjustbox}
\usepackage{amsfonts}
\usepackage{amssymb, amscd}
\usepackage{graphicx}
\usepackage{mathrsfs}
\usepackage{colortbl}
\usepackage{slashed}
\usepackage{url}
\usepackage{pdfpages}
\usepackage{braket}
\usepackage{sidecap}
\usepackage{arydshln}
\usepackage[font=small]{caption}
\usepackage{mathtools}
\usepackage{hyperref}
\usepackage{tikz}
\usepackage[compat=1.1.0]{tikz-feynman}
\usetikzlibrary{arrows,decorations.markings}
\usepackage{subcaption}
\numberwithin{equation}{section}

\usepackage{tikz}
\usepackage[compat=1.1.0]{tikz-feynman}

\newcommand{\Gprop}{
\begin{tikzpicture}[baseline]
\begin{feynman}
\vertex  [label=90:$\scriptstyle{\mu\nu}$](a);
\vertex [right=1.5cm of a, label=90:$\scriptstyle{\rho\sigma}$] (b);
\diagram* {
(a) -- [gluon, edge label=$k$] (b)
}; 
\end{feynman}
\end{tikzpicture}
}

\newcommand{\Gcub}{
\begin{tikzpicture}[baseline]
\begin{feynman}
\vertex (cc);
\vertex [above left=1.4cm of cc, label=90:$\scriptstyle{\alpha_1\beta_1}$] (a);
\vertex [below left=1.4cm of cc, label=270:$\scriptstyle{\alpha_2\beta_2}$] (b);
\vertex [right=1.2 cm of cc, label=90:$\scriptstyle{\alpha_3\beta_3}$] (c);
\diagram* {
(a) -- [gluon, momentum={[arrow shorten=0.3pt]$p_1$}] (cc),
(b) -- [gluon,  momentum={[arrow shorten=0.3pt]$p_2$}] (cc),
(c) -- [gluon,  momentum={[arrow shorten=0.3pt]$p_3$}] (cc),
}; 
\end{feynman}
\end{tikzpicture}
}

\newcommand{\Mdiag}{
\begin{tikzpicture}[baseline]
\begin{feynman}
\vertex [ dot] (a) {};
\vertex [below=2cm of a, dot] (b) {};
\vertex [right=1cm of a] (w);
\vertex [below=2cm of w] (u);
\vertex [right=2cm of a,  dot] (a') {};
\vertex [below=2cm of a', dot] (b') {};
\diagram* {
(b) -- [gluon] (a),
(a) -- [gluon, out=315, in=225, momentum={[arrow shorten=0.3pt]$k$}] (a'),
(b') -- [gluon] (a'),
(w) -- [scalar] (u)
}; 
\end{feynman}
\end{tikzpicture}
}

\newcommand{\Ndiag}{
\begin{tikzpicture}[baseline]
\begin{feynman}
\vertex [ dot] (a) {};
\vertex [below=2cm of a, dot] (b) {};
\vertex [right=2cm of a, dot] (a') {};
\vertex [below=2cm of a',  dot] (b') {};
\vertex [right=1cm of b] (w);
\vertex [right=1cm of a] (u);
\diagram* {
(b) -- [gluon ] (a),
(a) -- [gluon, momentum={[arrow shorten=0.4pt]$k$}] (b'),
(b') -- [gluon] (a'),
(u) -- [scalar] (w) 
}; 
\end{feynman}
\end{tikzpicture}
}

\newcommand{\IYdiag}{
\begin{tikzpicture}[baseline]
\begin{feynman}
\vertex [ dot] (a) {};
\vertex [below=2cm of a, dot] (b) {};
\vertex [right=2cm of b, dot] (b') {};
\vertex [above=1cm of b'] (c');
\vertex [above=2cm of b', dot] (a') {};
\vertex [right=1cm of b] (w);
\vertex [right=1cm of a] (u);
\diagram* {
(b) -- [gluon] (a),
(a) -- [gluon, momentum={[arrow shorten=0.4pt]$k$}] (c'),
(b') -- [gluon] (c') -- [gluon] (a'),
(u) -- [scalar] (w) 
}; 
\end{feynman}
\end{tikzpicture}} 

\newcommand{\Hdiag}{
\begin{tikzpicture}[baseline]
\begin{feynman}
\vertex [dot] (a) {};
\vertex [below=2cm of a, dot] (b) {};
\vertex [below=1cm of a] (c);
\vertex [right=2cm of c] (c');
\vertex [above=1cm of c', dot] (a') {};
\vertex [below=2cm of a', dot] (b') {};
\vertex [right=1cm of b] (w);
\vertex [right=1cm of a] (u);
\diagram* {
(b) -- [gluon] (c) -- [gluon] (a),
(b') -- [gluon] (c') -- [gluon] (a'),
(c) -- [gluon, momentum={[arrow shorten=0.4pt]$k$}] (c'),
(w) -- [scalar] (u)
}; 
\end{feynman}
\end{tikzpicture}}

\newcommand{\Azerodiag}{
\begin{tikzpicture}[baseline]
\begin{feynman}
\vertex [dot, label=180:$\scriptstyle{G^{1/2}}$] (a) {};
\vertex [below=1cm of a] (c);
\vertex [right=1cm of c] (d);
\vertex [below=2cm of a] (b);
\diagram* {
(a) -- [gluon] (d),
(a) -- [draw=none] (b)
}; 
\end{feynman}
\end{tikzpicture}} 

\newcommand{\Azerotwodiag}{
\begin{tikzpicture}[baseline]
\begin{feynman}
\vertex  (a);
\vertex [below=1cm of a] (c);
\vertex [right=1cm of c] (d);
\vertex [below=2cm of a, dot, , label=180:$\scriptstyle{G^{1/2}}$] (b) {};
\diagram* {
(b) -- [gluon] (d),
(b) -- [draw=none] (b)
}; 
\end{feynman}
\end{tikzpicture}} 

\newcommand{\Aonediag}{
\begin{tikzpicture}[baseline]
\begin{feynman}
\vertex [ dot, label=180:$\scriptstyle{G}$] (a) {};
\vertex [below=1cm of a] (c);
\vertex [right=1cm of c] (d);
\vertex [below=2cm of a, dot, label=180:$\scriptstyle{G^{1/2}}$] (b) {};
\diagram* {
(b) -- [gluon] (a),
(a) -- [gluon] (d)
}; 
\end{feynman}
\end{tikzpicture}} 

\newcommand{\Atwodiag}{
\begin{tikzpicture}[baseline]
\begin{feynman}
\vertex [dot, label=180:$\scriptstyle{G^{1/2}}$] (a) {};
\vertex [below=1cm of a] (c);
\vertex [right=1cm of c] (d);
\vertex [below=2cm of a,  dot, label=180:$\scriptstyle{G}$] (b) {};
\diagram* {
(a) -- [gluon] (b),
(b) -- [gluon] (d)
}; 
\end{feynman}
\end{tikzpicture}} 

\newcommand{\Athreediag}{
\begin{tikzpicture}[baseline]
\begin{feynman}
\vertex [dot, label=180:$\scriptstyle{G^{1/2}}$] (a) {};
\vertex [below=2cm of a, dot, label=180:$\scriptstyle{G^{1/2}}$] (b) {};
\vertex [below=1cm of a, label=180:$\scriptstyle{G^{1/2}}$] (c);
\vertex [right=1cm of c] (d);
\diagram* {
(b) -- [gluon] (c) -- [gluon] (a),
(c) -- [gluon] (d)
}; 
\end{feynman}
\end{tikzpicture}}

\newcommand{\Aonediagl}{
\begin{tikzpicture}[baseline]
\begin{feynman}
\vertex [ dot] (a) {};
\vertex [below=1cm of a] (c);
\vertex [right=1cm of c] (d);
\vertex [below=2cm of a, dot] (b) {};
\diagram* {
(b) -- [gluon] (a),
(a) -- [gluon, edge label=$k$] (d)
}; 
\end{feynman}
\end{tikzpicture}} 

\newcommand{\Atwodiagl}{
\begin{tikzpicture}[baseline]
\begin{feynman}
\vertex [dot] (a) {};
\vertex [below=1cm of a] (c);
\vertex [right=1cm of c] (d);
\vertex [below=2cm of a,  dot] (b) {};
\diagram* {
(a) -- [gluon] (b),
(b) -- [gluon, edge label=$k$] (d)
}; 
\end{feynman}
\end{tikzpicture}} 

\newcommand{\Athreediagl}{
\begin{tikzpicture}[baseline]
\begin{feynman}
\vertex [dot] (a) {};
\vertex [below=2cm of a, dot] (b) {};
\vertex [below=1cm of a] (c);
\vertex [right=1cm of c] (d);
\diagram* {
(b) -- [gluon] (c) -- [gluon] (a),
(c) -- [gluon, edge label=$k$] (d)
}; 
\end{feynman}
\end{tikzpicture}}

\newcommand{\Asqdiag}{
\begin{tikzpicture}[baseline]
\begin{feynman}
\vertex [blob, minimum size=0.7cm, fill=lightgray] (c) {${\cal A}_\lambda$};
\vertex [right=1.3cm of c] (d);
\diagram* {
(c) -- [gluon] (d)
}; 
\end{feynman}
\end{tikzpicture}}

\newcommand{\Asqdiaglab}{
\begin{tikzpicture}[baseline]
\begin{feynman}
\vertex [blob, minimum size=0.7cm, fill=lightgray] (c) {${\cal A}_\lambda$};
\vertex [right=1.3cm of c] (d);
\diagram* {
(c) -- [gluon, edge label=$k$] (d)
}; 
\end{feynman}
\end{tikzpicture}}

\newcommand{\Vtovdiag}{
\begin{tikzpicture}[baseline]
\begin{feynman}
\vertex [blob, minimum size=0.7cm, fill=lightgray] (c) {$\tilde{T}$};
\vertex [blob, right=2cm of c, minimum size=0.7cm, fill=lightgray] (d) {$\tilde{T}^*$};
\vertex [right=1cm of c] (cc);
\vertex [above=0.5cm of cc] (w);
\vertex [below=0.5cm of cc] (u);
\diagram* {
(c) -- [gluon, momentum={[arrow shorten=0.3pt]$k$}] (d),
(w) -- [scalar] (u);
}; 
\end{feynman}
\end{tikzpicture}}

\newcommand{\Mcutlabel}{
\begin{tikzpicture}[baseline]
\begin{feynman}
\vertex (c);
\vertex [left=0.8cm of c] (c');
\vertex [above=0.6cm of c'] (a) {1} ;
\vertex [below=0.6cm of c'] (b) {2};
\vertex [right=0.8cm of a] (d);
\vertex [right=2.2cm of d] (e);
\vertex [right=0.8cm of e] (f);
\vertex [right=0.8cm of b] (d') ;
\vertex [right=2.2cm of d'] (e');
\vertex [right=0.8cm of e'] (f');
\vertex [right=1.1cm of d] (u);
\vertex [above=1.2cm of u] (w); 
\vertex [below=1.8cm of u] (w'); 
\diagram* {
(a) -- [plain, line width=2pt] (d) -- [plain, line width=2pt, edge label=$ (2 u_1 \cdot \ell_1)^2$] (e)-- [plain, line width=2pt] (f) ,
(b) -- [plain, line width=2pt] (d') -- [plain, line width=2pt, edge label'=$2 u_2 \cdot \ell_2$] (e')-- [plain, line width=2pt] (f') ,
(d) -- [plain, edge label'=$(\ell_2 - q)^2$] (d'),
(e) -- [plain, edge label=$\ell_2^2$] (e'),
(d) -- [plain, half left, edge label=$(\ell_1 + \ell_2 - q)^2$] (e),
(w) -- [scalar] (w')
}; 
\end{feynman}
\end{tikzpicture}
}

\newcommand{\Msumcut}{
\begin{tikzpicture}[baseline]
\begin{feynman}
\vertex (c);
\vertex [left=0.8cm of c] (c');
\vertex [above=0.6cm of c'] (a);
\vertex [below=0.6cm of c'] (b);
\vertex [right=0.8cm of a] (d);
\vertex [right=1cm of d] (e);
\vertex [right=0.8cm of e] (f);
\vertex [right=0.8cm of b] (d');
\vertex [right=1cm of d'] (e');
\vertex [right=0.8cm of e'] (f');
\vertex [left=0.4cm of d] (u);
\vertex [above=0.7cm of u] (w); 
\vertex [below=1.5cm of u] (w'); 
\diagram* {
(a) -- [plain, line width=2pt] (f),
(b) -- [plain, line width=2pt] (f'),
(d) -- [plain] (d'),
(e) -- [plain] (e'),
(d) -- [plain, half left] (e),
(w) -- [scalar] (w')
}; 
\end{feynman}
\end{tikzpicture} +\begin{tikzpicture}[baseline]
\begin{feynman}
\vertex (c);
\vertex [left=0.8cm of c] (c');
\vertex [above=0.6cm of c'] (a);
\vertex [below=0.6cm of c'] (b);
\vertex [right=0.8cm of a] (d);
\vertex [right=1cm of d] (e);
\vertex [right=0.8cm of e] (f);
\vertex [right=0.8cm of b] (d');
\vertex [right=1cm of d'] (e');
\vertex [right=0.8cm of e'] (f');
\vertex [right=0.5cm of d] (u);
\vertex [above=0.7cm of u] (w); 
\vertex [below=1.5cm of u] (w'); 
\diagram* {
(a) -- [plain, line width=2pt] (f),
(b) -- [plain, line width=2pt] (f'),
(d) -- [plain] (d'),
(e) -- [plain] (e'),
(d) -- [plain, half left] (e),
(w) -- [scalar] (w')
}; 
\end{feynman}
\end{tikzpicture} + \begin{tikzpicture}[baseline]
\begin{feynman}
\vertex (c);
\vertex [left=0.8cm of c] (c');
\vertex [above=0.6cm of c'] (a);
\vertex [below=0.6cm of c'] (b);
\vertex [right=0.8cm of a] (d);
\vertex [right=1cm of d] (e);
\vertex [right=0.8cm of e] (f);
\vertex [right=0.8cm of b] (d');
\vertex [right=1cm of d'] (e');
\vertex [right=0.8cm of e'] (f');
\vertex [right=0.4cm of e] (u);
\vertex [above=0.7cm of u] (w); 
\vertex [below=1.5cm of u] (w'); 
\diagram* {
(a) -- [plain, line width=2pt] (f),
(b) -- [plain, line width=2pt] (f'),
(d) -- [plain] (d'),
(e) -- [plain] (e'),
(d) -- [plain, half left] (e),
(w) -- [scalar] (w')
}; 
\end{feynman}
\end{tikzpicture} 
}

\newcommand{\VIcut}{
\begin{tikzpicture}[baseline]
\begin{feynman}
\vertex (c);
\vertex [left=0.6cm of c] (c');
\vertex [above=0.5cm of c'] (a);
\vertex [below=0.3cm of c'] (b);
\vertex [right=0.4cm of a] (d);
\vertex [right=0.4cm of d] (e);
\vertex [right=0.3cm of e] (g);
\vertex [right=0.4cm of g] (f);
\vertex [right=0.4cm of b] (d');
\vertex [right=0.7cm of d'] (e');
\vertex [right=0.4cm of e'] (f');
\vertex [right=0.15cm of e] (u);
\vertex [above=0.1cm of u] (w); 
\vertex [below=1cm of u] (w'); 
\diagram* {
(a) -- [plain, line width=2pt] (f),
(b) -- [plain, line width=2pt] (f'),
(d) -- [plain] (d') -- [plain] (e),
(g) -- [plain] (e'), 
(w) -- [scalar] (w')
}; 
\end{feynman}
\end{tikzpicture}
}

\newcommand{\Tmnloop}{
\begin{tikzpicture}[baseline]
\begin{feynman}
\vertex [blob, minimum size=0.7cm, fill=lightgray] (c) {};
\vertex [right=1.5cm of c, label=90:$\scriptstyle{\mu\nu}$] (d);
\diagram* {
(c) -- [gluon, edge label=$k$] (d)
}; 
\end{feynman}
\end{tikzpicture}}

\newcommand{\FRulezero}{
\begin{tikzpicture}[baseline]
\begin{feynman}
\vertex [dot, label=90:$\tau_a$] (a) {};
\vertex [right=0.8cm of a] (b);
\diagram* {
(a) -- [gluon, momentum={[arrow shorten=0.3pt]$k$}, insertion=1] (b),
}; 
\end{feynman}
\end{tikzpicture}}

\newcommand{\FRuleone}{
\begin{tikzpicture}[baseline]
\begin{feynman}
\vertex [ dot, label=90:$\tau_a$] (a) {};
\vertex [below=1.2cm of a, dot] (c) {};
\vertex [below right=0.8cm of a] (b);
\diagram* {
(c) -- [gluon] (a),
(a) -- [gluon, momentum={[arrow shorten=0.3pt]$k$}, insertion=1] (b)
}; 
\end{feynman}
\end{tikzpicture}
}

\newcommand{\Azerodiaglab}{
\begin{tikzpicture}[baseline]
\begin{feynman}
\vertex [dot, label=180:$1$] (a) {};
\vertex [below=1cm of a] (c);
\vertex [right=1cm of c] (d);
\vertex [below=2cm of a] (b);
\diagram* {
(a) -- [gluon, momentum={[arrow shorten=0.3pt]$k$}] (d)
}; 
\end{feynman}
\end{tikzpicture}}

\newcommand{\Aonediaglab}{
\begin{tikzpicture}[baseline]
\begin{feynman}
\vertex [ dot, label=180:$\tau_1$] (a) {};
\vertex [below=1cm of a] (c);
\vertex [right=1cm of c] (d);
\vertex [below=1.5cm of a, dot] (b) {};
\diagram* {
(b) -- [gluon] (a),
(a) -- [gluon, momentum={[arrow shorten=0.3pt]$k$}] (d)
}; 
\end{feynman}
\end{tikzpicture}}

\newcommand{\Athreediaglab}{
\begin{tikzpicture}[baseline]
\begin{feynman}
\vertex [dot] (a) {};
\vertex [below=2cm of a, dot] (b) {};
\vertex [below=1cm of a] (c);
\vertex [right=1cm of c] (d);
\diagram* {
(b) -- [gluon, momentum={[arrow shorten=0.3pt]$q_2$}] (c) -- [gluon, reversed momentum={[arrow shorten=0.3pt]$q_1$}] (a),
(c) -- [gluon, momentum={[arrow shorten=0.3pt]$k$}] (d)
}; 
\end{feynman}
\end{tikzpicture}}

\newcommand{\SIone}{
\begin{tikzpicture}[baseline]
\begin{feynman}
\vertex (c);
\vertex [left=1cm of c] (c');
\vertex [above=0.6cm of c'] (a) {1};
\vertex [below=0.6cm of c'] (b) {2};
\vertex [right=1cm of a] (d);
\vertex [right=1.4cm of d] (e);
\vertex [right=1cm of e] (f);
\vertex [right=1cm of b] (d');
\vertex [right=1.4cm of d'] (e');
\vertex [right=1cm of e'] (f');
\diagram* {
(a) -- [plain, line width=2pt] (f),
(b) -- [plain, line width=2pt] (f'),
(d) -- [plain] (d'),
(e) -- [plain] (e'),
(d) -- [plain, half left] (e)
}; 
\end{feynman}
\end{tikzpicture}} 

\newcommand{\SItwo}{
\begin{tikzpicture}[baseline]
\begin{feynman}
\vertex (c);
\vertex [left=1cm of c] (c');
\vertex [above=0.6cm of c'] (a) {1};
\vertex [below=0.6cm of c'] (b) {2};
\vertex [right=1cm of a] (d);
\vertex [right=1.4cm of d] (e);
\vertex [right=1cm of e] (f);
\vertex [right=1cm of b] (d');
\vertex [right=1.4cm of d'] (e');
\vertex [right=1cm of e'] (f'); 
\diagram* {
(a) -- [plain, line width=2pt] (f),
(b) -- [plain, line width=2pt] (f'),
(d) -- [plain] (d') -- [plain] (e),
(e) -- [plain] (e')
}; 
\end{feynman}
\end{tikzpicture}}

\newcommand{\SIthree}{
\begin{tikzpicture}[baseline]
\begin{feynman}
\vertex (c);
\vertex [left=1cm of c] (c');
\vertex [above=0.6cm of c'] (a) {1};
\vertex [below=0.6cm of c'] (b) {2};
\vertex [right=1cm of a] (d);
\vertex [right=0.9cm of d] (e);
\vertex [right=0.5cm of e] (g);
\vertex [right=1cm of g] (f);
\vertex [right=1cm of b] (d');
\vertex [right=1.4cm of d'] (e');
\vertex [right=1cm of e'] (f');
\diagram* {
(a) -- [plain, line width=2pt] (f),
(b) -- [plain, line width=2pt] (f'),
(d) -- [plain] (d') -- [plain] (e),
(g) -- [plain] (e')
}; 
\end{feynman}
\end{tikzpicture}} 

\newcommand{\SIfour}{
\begin{tikzpicture}[baseline]
\begin{feynman}
\vertex (c);
\vertex [left=1cm of c] (c');
\vertex [above=0.6cm of c'] (a) {1};
\vertex [below=0.6cm of c'] (b) {2};
\vertex [right=1cm of a] (d);
\vertex [right=1.4cm of d] (e);
\vertex [right=1cm of e] (f);
\vertex [right=1cm of b] (d');
\vertex [right=1.4cm of d'] (e');
\vertex [right=1cm of e'] (f');
\vertex [below=1cm of d] (h);
\vertex [below=1cm of e] (h');
\diagram* {
(a) -- [plain, line width=2pt] (f),
(b) -- [plain, line width=2pt] (f'),
(d) -- [plain] (d'),
(e) -- [plain] (e'),
(h) -- [plain] (h')
}; 
\end{feynman}
\end{tikzpicture}}

\hypersetup{
    colorlinks=true,
    linkcolor=ceruleanblue,
    filecolor=ceruleanblue,      
    urlcolor=ceruleanblue,
    citecolor=ceruleanblue,
}

\definecolor{myblue}{RGB}{174, 198, 219}
\definecolor{myred}{RGB}{157,31,68}
\definecolor{ceruleanblue}{rgb}{0.0, 0.2, 0.6}

\newcommand{\be}{\begin{equation}}      
\newcommand{\ee}{\end{equation}}

\newcommand{\mpl}{m_{\text{Pl}}}

\newcommand{\Ord}[1]{\mathit{O}\left(#1\right)}

\newcommand{\vv}{{\bf v}}
\newcommand{\vb}{{\bf b}}
\newcommand{\vk}{{\bf k}}
\newcommand{\vq}{{\bf q}}

\newcommand{\vn}{{\bf n}}
\newcommand{\ve}{{\bf e}}

\def\dd{\delta\!\!\!{}^-\!}
\def\ddp{\delta\!\!\!{}^-\!{}_+}
\newcommand{\uu}{{\cal U}}
\newcommand{\Imm}[1]{\mathrm{Im}\left(#1\right)}

\newcommand{\Real}{\mathbb{R}}
\newcommand{\firstc}{\tikz{\draw[line width=0.5pt] (0,-5pt) -- (0,0pt) -- (3pt,-2pt) ;}}
\newcommand{\secondc}{\tikz{\draw[line width=0.5pt] (0,0pt) -- (0,-5pt) -- (3pt,-2pt) ;}}

\DeclareMathOperator{\arcsinh}{arcsinh}

\abstract{We compute the four-momentum radiated  during the scattering of two spinless bodies, at leading order in the Newton constant $G$ and at all orders in the velocities, using the Effective Field Theory worldline approach. 
Following \cite{Mougiakakos:2021ckm}, we derive the conserved stress-energy tensor linearly coupled to gravity generated by localized sources,  at leading and next-to-leading order in $G$,  and from that the classical probability amplitude of graviton emission.   The total emitted momentum is obtained by phase-space integration of the graviton momentum weighted by the modulo squared of the radiation amplitude. We recast this as a two-loop integral that we solve using techniques borrowed from particle physics, such as reverse unitarity, reduction to master integrals by integration-by-parts identities and canonical differential equations. 
The emitted momentum agrees with  recent results obtained by other methods. Our approach provides an alternative way of directly computing radiated observables in the post-Minkowskian expansion without going through the classical limit  of scattering amplitudes.}

\begin{document}

\maketitle
\flushbottom

\section{Introduction}

Gravitational waves from  black hole and neutron star binaries will provide an unprecedented source of information about astrophysics, cosmology and fundamental physics. 
This will be possible thanks to an increase in sensitivity of future  detectors, such as LISA \cite{LISA:2017pwj}, Einstein Telescope \cite{Punturo:2010zz}  and Cosmic Explorer \cite{Reitze:2019iox}, which will require improving the   modeling of the relativistic two-body dynamics over the current state-of-the-art \cite{Purrer:2019jcp}. 

Currently,  waveform templates are  modeled using semi-analytical approaches such as the effective-one-body formalism \cite{Buonanno:2000ef}. 
A crucial ingredient of this approach is the energy map between the two-body and the effective one-body systems, originally established using the post-Newtonian (PN) approximation \cite{Blanchet:2013haa,Blanchet:2018hut}. This consists in expanding for small gravitational potential $G m/r$, with $m$ the typical mass of the two objects and $r$ their relative distance, and small relative velocity between the two bodies $v$.  In \cite{Damour:2016gwp,Damour:2017zjx}, it has been suggested that this mapping can be improved by using the post-Minkowskian (PM) approximation method, i.e.~by expanding in the gravitational constant $G$ without assuming a small   velocity (see \cite{Bertotti:1956pxu,Bertotti:1960wuq,Havas:1962zz,Westpfahl:1979gu,Portilla:1980uz,Bel:1981be,Westpfahl:1985tsl} for early works on the PM approximation). This has sparked fervent activity in the application of PM methods to the two-body relativistic gravitational dynamics.

While the PN expansion is natural for gravitationally bound systems---after all, the gravitational potential and velocity are related by the virial theorem---the PM expansion applies naturally to unbound systems, such as the scattering of two massive black holes.  
Information about their dynamics can be extracted by modeling the scattering black holes as a quantum mechanical system of two scattering particles. This  can be studied on the basis of the long-time well-established quantum field theory description of gravity  \cite{Veltman:1975vx,DeWitt:1967yk,DeWitt:1967ub,DeWitt:1967uc,Donoghue:1994dn,Bjerrum-Bohr:2002gqz,Iwasaki:1971iy,Iwasaki:1971vb}, 
by   taking its appropriate classical limit. For a bound binary system, the classical limit consists in the orbital angular momentum $L = m v r$ much larger than $\hbar$ or, in natural units, $L \gg 1$ \cite{Goldberger:2004jt}. For a scattering process $L = p b$, where $p$ is the asymptotic center-of-mass  momentum of the particles and $b$ the impact parameter, so that the classical limit is obtained for $q \sim 1/b \ll p$, where $q$ is the momentum exchanged in the scattering process.
Such a small $q$ expansion is analogous to the so-called soft expansion familiar from the method of regions \cite{Beneke:1997zp}.

Following this program, scattering amplitude techniques, such as the double copy \cite{Bern:2008qj,Bern:2010ue,Bern:2019prr,Brandhuber:2021kpo}, generalized unitarity \cite{Bern:1994zx,Bern:1994cg,Britto:2004nc} and   effective field theory (EFT) matching \cite{Neill:2013wsa,Bjerrum-Bohr:2013bxa,Luna:2017dtq,Bjerrum-Bohr:2018xdl,Cheung:2018wkq,Cristofoli:2019neg,Cristofoli:2020uzm}, have recently been used to study the two-body conservative dynamics at increasing PM orders \cite{Bern:2019nnu,Bern:2019crd,Cheung:2020gyp,Bern:2021dqo}, and as well including tidal effects \cite{Bern:2020uwk,Cheung:2020sdj,AccettulliHuber:2020oou,Haddad:2020que,Aoude:2020onz,Cheung:2020gbf} and spins \cite{Arkani-Hamed:2017jhn,Chung:2018kqs,Vines:2018gqi,Bern:2020buy,Guevara:2018wpp}.  Interestingly,  observables of bound and unbound systems have been shown to be  related, in some cases,  via analytic continuation \cite{Kalin:2019rwq,Kalin:2019inp}. 

Scattering bodies are accompanied by emission of  gravitational  Bremsstrahlung radiation \cite{Peters:1970mx,Thorne:1975aa,Crowley:1977us,Kovacs:1977uw,Kovacs:1978eu,Turner:1978zz,Westpfahl:1985tsl}, which is the unbound analog of the gravitational waves emitted by inspiral binaries and is suppressed by three powers of $G$. 
Although radiation reaction effects were  thoroughly investigated in the past in the Regge limit (i.e., when the center-of-mass energy is much larger than the momentum transfer) \cite{Amati:1990xe,DiVecchia:2019myk,DiVecchia:2019kta,Bern:2020gjj} or in association to the loss of angular momentum in the collision \cite{DiVecchia:2020ymx,Damour:2020tta,DiVecchia:2021ndb,Bjerrum-Bohr:2021vuf,Bjerrum-Bohr:2021din,Brandhuber:2021eyq},  
the full leading-order emitted momentum  has been  obtained only very recently in \cite{Herrmann:2021lqe,Herrmann:2021tct} via the formalism of \cite{Kosower:2018adc}, which derives classical observables from quantum scattering  (see also  \cite{Maybee:2019jus,delaCruz:2020bbn}  for extensions of the formalism of \cite{Kosower:2018adc} to spin and classical observables in Yang-Mills theories), and in \cite{DiVecchia:2021bdo} using  the eikonal approach to classical gravitational scattering. These calculations require evaluating the classical limit of relevant two-loop Feynman integrals, that can be solved by combining different techniques borrowed from particle physics, as shown in 
\cite{Parra-Martinez:2020dzs}, namely reduction to master integrals   by Integration-by-Parts (IBP) identities \cite{Tkachov:1981wb,Chetyrkin:1981qh, Smirnov:2012gma} and   differential equations \cite{Kotikov:1990kg,Bern:1992em,Gehrmann:1999as,Henn:2013pwa} to solve the latter, using the near-static regime as initial conditions.

In this paper we focus on the computation of the   emitted momentum at leading PM order using an EFT worldline approach inspired by  Non-Relativistic-General-Relativity (NRGR) \cite{Goldberger:2004jt}. 
In the traditional  NRGR approach (see \cite{Goldberger:2007hy,Foffa:2013qca,Rothstein:2014sra,Porto:2016pyg,Levi:2018nxp} for reviews) one builds an EFT of classically radiating gravitons by exploiting the separation of the  three relevant scales in the system: the size of the inspiraling bodies, the orbital radius and   the wavelength of the emitted gravitational radiation. 
The  potential-graviton propagators are expanded in the small velocity limit in a PN expansion and then integrated out. The bodies are treated as static background sources for the graviton dynamics, i.e.,  their  recoil due to graviton interaction  is neglected  because suppressed by  $q/p \sim 1/L \ll 1$, where $q$ is the momentum of the exchanged graviton and $p$ the typical momentum of the bodies. This 
approximation amounts to impose that the system is classical from the onset, dispensing one from the  (sometimes tedius) $\hbar$ counting.
Quantum corrections, described by graviton loops, are suppressed by powers of $1/L$ relative to  tree diagrams and can be ignored.

The  worldline approach has been extended to the PM approximation \cite{Goldberger:2016iau,Goldberger:2017vcg,Shen:2018ebu,Kalin:2020mvi,Mogull:2020sak,Loebbert:2020aos}. In this scheme, one expands the body trajectories around rectilinear motion, each order in the expansion carrying an additional power of $G$. Combined with the  powerful methods of IBP reduction and differential equations to solve the integrals involved in the calculations, this approach
has allowed to quickly reach most of the state-of-the-art results achieved by scattering-amplitude techniques \cite{Kalin:2020fhe,Kalin:2020lmz,Liu:2021zxr,Cho:2021mqw,Dlapa:2021npj}.

In \cite{Mougiakakos:2021ckm} (see also \cite{Jakobsen:2021smu}) we  applied this approach to compute the  conserved stress-energy tensor linearly coupled to gravity generated by the two bodies  at leading and next-to-leading   order in $G$  and, from that, the classical probability amplitude of graviton emission. The radiated four-momentum is  given by a phase-space integral of the graviton momentum, weighted by the modulo squared of the radiation amplitude.     At leading order, the radiation amplitude is just a static piece that does not contribute to the emitted energy.
The next-to-leading   amplitude, instead, contributes at leading order. It is given by an  integral in the graviton momentum  exchanged between the two bodies, but we were unable to perform this integral and write it  in terms of known functions.  (We note, in passing, that the  Fourier transform of the amplitude, which is simply the waveform in time domain, can instead be performed, leading to the  expression originally computed in \cite{Kovacs:1977uw, Kovacs:1978eu} (see also \cite{Jakobsen:2021smu}).) We could express it in compact form as a one-dimensional integral over a Feynman parameter involving Bessel functions. Using this, we  recovered the leading-order radiated angular momentum \cite{Damour:2020tta} and, upon expansion of the integrand in $v$, the total four-momentum radiated into gravitational waves  up to  order $v^8$, finding agreement with \cite{Herrmann:2021lqe}.  

We review these developments in Sec.~\ref{Sec2}, spelling out details missing in \cite{Mougiakakos:2021ckm}. In particular, we lay out the formalism of the EFT worldline approach for the PM calculation and  introduce the Feynman rules. We then 
compute the stress-energy tensor and  the classical amplitude of graviton emission at leading- and next-to-leading order. Appendix~\ref{AppA} contains the expression of the conserved stress-energy tensor in arbitrary spacetime dimensions. Contrarily to the more compact expression  used in the main text, this stress-energy tensor is conserved both on-shell and off-shell. Appendix \ref{AppB} details the treatment of the Feynman integrals involved in the amplitude computation.

In Sec.~\ref{Sec3} we  bypass  the problem of not having a solution for the amplitude by rewriting the 
phase-space integral of the four-momentum as a (cut) two-loop integral. Using reverse unitarity  \cite{Anastasiou:2002yz,Anastasiou:2002qz,Anastasiou:2003yy,Anastasiou:2015yha}, we treat the phase-space delta function  as a cut propagator. Following  \cite{Parra-Martinez:2020dzs}, in Sec.~\ref{Sec4} we solve this integral using techniques developed in QCD and high-energy physics. In particular, we organize our calculations in terms of four topologies that come out naturally from our Feynman rules for the gravitons and we solve each topology, one by one.\footnote{The details of these calculations can be found in the ancillary files attached to the \texttt{arXiv} submission of this article as \texttt{Mathematica} notebooks.}
This shows that these techniques can be  applied to the wordline approach to derive classical observables, without going through the classical limit of scattering amplitudes.
Appendix \ref{AppC} contains a thorough discussion on the boundary conditions necessary to solve the ordinary differential equations involved by the master integrals and details how to compute their solutions. 
Finally, we conclude in Sec.~\ref{Sec5}.

We use the mostly minus convention for the signature of the metric, the following notation for the integration symbol,
\be
\int_{q} \equiv \int \frac{d^4 q }{(2 \pi)^4} \;,
\ee
and the definitions $\dd^{(n)} (x) \equiv (2 \pi)^n \delta^{(n)} (x)$ and $\dd_+(p^2) \equiv \dd (p^2) \theta(p^0)$.
Unless otherwise specified, we will use natural units with $\hbar = c= 1$ and define the  Planck mass as
\be
\label{mpl}
\mpl \equiv \frac{1}{\sqrt{32 \pi G}} \;.
\ee

\section{From the action to the emission amplitude}
\label{Sec2}

Following \cite{Mougiakakos:2021ckm}, we outline here the general set up of the  PM worldline approach and derive the stress-energy tensor and, finally, the amplitude of classical gravitational emission.

\subsection{Post-Minkowskian effective field theory}

We consider a system of two spinless massive bodies with masses $m_1$ and $m_2$, interacting via  gravity, described by the Einstein-Hilbert action.
As discussed in the introduction, we  rely on the separation between the relevant scales in the system. 
We will assume that the impact parameter $b$ involved in the collision between the two bodies is much larger than their typical size. We thus treat
the two bodies as point particles. Finite size effects can be incorporated systematically as higher-derivative operators along the worldline (see e.g.~\cite{Kalin:2020mvi,Kalin:2020lmz}).
The objects are considered as external (i.e.~non-propagating) sources of the gravitational field.
The resulting action, describing the dynamics of the system, is then \cite{Kalin:2020mvi} 
\be
\label{eq:actionone}
S = -2 \mpl^2\int\!\! d^4 x \sqrt{-g} R - \sum_a \frac{m_a}{2} \int\!\! d\tau_a \big[ g_{\mu\nu}(x_a)\, {\cal U}_a^\mu(\tau_a) {\cal U}_a^\nu (\tau_a) + 1 \big] \, ,
\ee
where, for each body $a=1, 2$, $\tau_a$ is the proper time,  ${\cal U}_a^\mu (\tau_a) = d x_a^\mu / d \tau_a$ is the four-velocity and $\mpl$ is defined in \eqref{mpl}. Note that for the above action we have used a Polyakov-like parametrization, which has the advantage of simplifying the coupling of matter with gravity \cite{Kalin:2020mvi, Galley:2013eba, Kuntz:2020gan}.

We want to compute the classical pseudo-stress-energy tensor $T^{\mu\nu}(x)$, defined as the linear terms sourcing the gravitational field in the effective action \cite{DeWitt:1967ub,Abbott:1981ke,Goldberger:2004jt}, i.e., 
\be
\label{bfea}
\Gamma [x_a, h_{\mu \nu} ] = - \frac{1}{2 \mpl} \int d^4 x T^{\mu \nu} (x) h_{\mu \nu} (x) \;,
\ee
where $h_{\mu\nu} (x) = \mpl (g_{\mu\nu} - \eta_{\mu\nu})$. This includes all contribution coming  from both the external sources, i.e. the point-particles, and the gravitational self-interaction. 

We can compute $T^{\mu\nu}$ via a matching procedure. In particular, we expand the action \eqref{eq:actionone}  for small $h_{\mu\nu}$ and use this to compute the one-point expectation value $\langle h_{\mu \nu} \rangle$, considering
all Feynman diagrams that involve one external graviton. We do the same using an effective action composed by the quadratic action of $h_{\mu \nu}$ plus an interaction term \eqref{bfea}. We can then find $T^{\mu \nu}$ by matching the two results. 
Pictorially, we can depict this procedure as follows
\begin{equation}
\label{eq:match}
\Tmnloop = \frac{1}{2 \mpl} P_{\mu\nu\rho\sigma} \frac{\tilde{T}^{\rho\sigma}(k)}{k^2}  \, ,
\end{equation}
where the left-hand side stands for all the possible Feynman diagrams with one external graviton. On the right-hand side we have denoted the Fourier transform by a tilde,
\be
\tilde X(k) = \int d^4 x X(x) e^{i k \cdot x} \;.
\ee 
This procedure can be done order by order in the perturbative expansion in $G$.

Once $T^{\mu\nu}$ is known, we can use it to compute the classical probability amplitude of emitting one graviton with helicity $\lambda$ and momentum $k^\mu$, defined by
\be
\label{amplitude}
i {\cal A}_\lambda( k) = -  \frac{i}{2 \mpl} \epsilon^{ * \lambda}_{\mu \nu} (\vk)  \tilde T^{\mu \nu}(k) \;,
\ee 
where $\epsilon_{\mu \nu}^\lambda (\vk)$ is the helicity-2 polarization tensor, with normalization $\epsilon^{* \lambda}_{\mu \nu} (\vk) \epsilon_{\lambda'}^{\mu \nu} (\vk) = \delta^{\lambda}_{ \lambda'}$. This 
is directly related to the asymptotic waveform (see e.g.~\cite{Maggiore:1900zz}) by
\be
\label{waveform}
h_{\mu \nu} (x) = - \frac1{4 \pi r} \sum_{\lambda = \pm 2} \int \frac{d k^0}{2 \pi} e^{-i k^0 u} \epsilon_{\mu \nu}^\lambda (\vk) {\cal A}_\lambda (k) |_{k^\mu = k^0 n^\mu} \;, 
\ee
where $r$ is a distance much larger than the interaction region, and 
\be
\label{eq:defn}
n^\mu = (1, \vn) \, ,
\ee
with $\vn$  the unitary vector pointing along the direction of propagation of the emitted graviton. Equation \eqref{amplitude} can also be used to compute radiated observables such as the radiated linear momentum, as we will see in Sec.~\ref{Sec:2loop_int}.

Let us introduce, then, the Feynman rules relevant for the computation of the leading-order and next-to-leading-order stress-energy tensor. As usual, for the gravitational sector of the action \eqref{eq:actionone} we need to introduce a gauge-fixing term in order to define a propagator for $h_{\mu\nu}(x)$. We choose to work in the so-called de Donder gauge, which consists in adding the following gauge-fixing action,
\begin{equation}
S_{\text{GF}} = \int\!d^4x\, \left(\partial^\rho h_{\rho \mu} - \frac{1}{2}\partial_{\mu} h \right)\left(\partial_\sigma h^{\sigma \mu} - \frac{1}{2}\partial^{\mu} h \right) \, ,
\end{equation}
where $h \equiv h_{\mu\nu}\eta^{\mu\nu}$. Since we want only classical contributions, in \eqref{eq:match} we exclude diagrams involving closed graviton loops, as they are purely quantum  \cite{Goldberger:2004jt, Porto:2016pyg}. For the same reason, ghost terms are unnecessary. 

From the expansion of the gravitational action at quadratic order, we then define the usual graviton free propagator in de Donder gauge, for $h_{\mu\nu}$, i.e.,
\begin{align}
\label{eq:propagator}
\Gprop = \frac{i}{k^2} P_{\mu\nu\rho\sigma} \, , & & P_{\mu\nu\rho\sigma} = \eta_{\mu (\rho} \eta_{\sigma) \nu} - \frac{1}{2}\eta_{\mu\nu}\eta_{\rho\sigma} \, .
\end{align}
As usual, one must specify the contour of integration in the complex plane $k^0$, and this can be done by choosing the suitable $i 0^+$ prescription in the denominator. Since we want to take into account only outgoing graviton, one should impose retarded boundary conditions, i.e. $\left[(k^0 + i0^+)^2 - |\vk|^2 \right]^{-1}$. However, this is not relevant for the current computation, because, as we shall see, all the integrated graviton momenta are off-shell, so that we do not need to specify the $i0^+$ prescription for these propagators. This will no longer be true at higher orders, where hereditary effects  contribute to the computation of $T^{\mu\nu}$ \cite{Goldberger:2009qd, Galley:2015kus}. 
Expanding the action at higher orders gives the self-interaction vertices. For the current computation  we will just need the cubic  vertex
\begin{align}
\label{eq:cubicV}
\Gcub = \frac{i}{\mpl} \dd^{(4)}(p_1+ p_2 + p_3) V_3^{\alpha_1 \beta_1 \alpha_2 \beta_2 \alpha_3 \beta_3} (p_1, p_2, p_3) \, ,
\end{align}
where we have expanded the action using the \texttt{Mathematica} packages \texttt{xTensor} and \texttt{xPert} \cite{Brizuela:2008ra, Martin-Garcia:xAct} to compute  $V_3^{\alpha_1 \beta_1 \alpha_2 \beta_2 \alpha_3 \beta_3}$. This is bilinear in the momenta, symmetric in $\alpha_a$ and $\beta_a$ for $a=1, 2, 3$ and symmetric in the exchange of $(p_1, \alpha_1 \beta_1)$, $(p_2, \alpha_2 \beta_2)$ and $(p_3, \alpha_3 \beta_3)$. We do not provide the explicit expression here because of its length.

Finally, we need to write down the Feynman rules coming from the interaction of gravity with the external sources. As one can see from eq.~\eqref{eq:actionone}, in principle we have  just one linear interaction vertex. However, in order to completely isolate the powers of $G$, we expand perturbatively  the worldline around straight motion \cite{Kalin:2020mvi, Kalin:2020fhe}, i.e. 
\begin{align}
\label{eq:exp_x}
x_a^\mu(\tau_a) & = b^\mu_a + u^\mu_a \tau_a + \delta^{(1)} x_a^\mu(\tau_a) +\dots \; ,\\
\label{eq:exp_u}
\uu_a^\mu (\tau_a) & = u_a^{\mu} + \delta^{(1)} u_a^\mu (\tau_a) +\dots \; .
\end{align}
Here $ u_a$ is the (constant) asymptotic incoming velocity and $b_a$ is the body displacement orthogonal to  it, $b_a \cdot u_a =0$. 
With this expansion, at leading order we obtain  the following Feynman rules, 
\begin{equation}
\label{eq:Fzero}
\FRulezero = -\frac{i m_a}{2\mpl} u^\mu_a u^\nu_a \int d\tau_a e^{i k \cdot (b_a+u_a \tau_a)}  \; ,
\end{equation}
where a bullet stands for the point particle and the cross attached to the wiggly line is there to remind us that there is no propagator attached to the straight worldline. At first order in $G$ we have
\be
\label{eq:Fone}
\raisebox{20pt}{\FRuleone} = -\frac{i m_a}{2\mpl} \int d\tau_a e^{i k \cdot (b_a+u_a \tau_a)} \left( 2 \delta^{(1)} u_a^{(\mu}(\tau_a) u_a^{\nu)} + i ( k \cdot  \delta^{(1)} x_a (\tau_a) ) u_a^\mu u_a^\nu \right)  \; .
\ee
For the computation of this paper we can stop at this order.

In order to compute the first-order deviations from straight trajectories,  $\delta^{(1)} x_a^\mu$ and $ \delta^{(1)} u_a^\mu$, one has to compute the effective action by integrating out  the graviton  from eq.~\eqref{eq:actionone}, i.e.
\begin{equation}
e^{i S_{\rm eff}[x_a]} = \int {\cal D} [h] e^{i(S + S_{\rm GF})} \, .
\end{equation}
Varying $S_{\rm eff}[x_a]$, one can then derive and solve the equations of motion for the two point particles. This procedure can be done perturbatively in the Newton constant $G$ as explained in \cite{Kalin:2020mvi}. In general, one must carefully include all the contributions  from both the potential and the radiation modes of $h_{\mu\nu}$. However,  the leading-order corrections  $\delta^{(1)} x_a^\mu$ and $ \delta^{(1)} u_a^\mu$ are determined by only potential gravitons. In particular, for the first order corrections we just need the following action in the de Donder gauge \cite{Kalin:2020mvi}
\begin{equation}
S^{(1)}_{\rm eff} = -\frac{m_1 m_2}{8 \mpl^2}\int d\tau_1 d\tau_2 \Big[2 \big( {\cal U}_1(\tau_1) \cdot {\cal U}_2(\tau_2)\big)^2 -{\cal U}^2_1(\tau_1) {\cal U}^2_2(\tau_2) \Big] \int_q \frac{e^{-i q \cdot ( x_1 (\tau_1) - x_2(\tau_2) ) }}{q^2} \, .
\end{equation}
One can then vary this action to find the equations of motion for the two worldlines, and then expand as in eqs. \eqref{eq:exp_x} and \eqref{eq:exp_u} to solve them perturbatively in $G$. For particle $1$ one eventually gets
\begin{align}
\label{deltau}
\delta^{(1)} u_1^\mu (\tau)= & \frac{m_2}{4 \mpl^2}\!\int_q \dd(q \cdot u_2) \frac{e^{-i q\cdot b - i q\cdot u_1 \tau}}{q^2} \left( \frac{2\gamma^2-1}{2}\frac{q^\mu}{q\cdot u_1 + i0^+}-2\gamma u_2^\mu +u_1^\mu \right) \;,  \\
\delta^{(1)} x_1^\mu (\tau)= & \frac{i m_2}{4 \mpl^2}\!\int_q \dd(q \cdot u_2) \frac{e^{-i q\cdot b - i q\cdot u_1 \tau}}{q^2 (q\cdot u_1 + i0^+)} \left( \frac{2\gamma^2-1}{2}\frac{q^\mu}{q\cdot u_1 + i0^+}-2\gamma u_2^\mu +u_1^\mu \right) \;,
\label{deltax}
\end{align}
where 
\be
\gamma \equiv u_1 \cdot u_2\;,
\ee
 and $b  \equiv b_1 -b_2$.
An analogous expression holds for particle 2. The $+i 0^+$ in the above equations ensures to recover  straight motion in the asymptotic past, i.e.~$\delta^{(1)} u_1^\mu (- \infty)=0$ and $\delta^{(1)} x_1^\mu (- \infty)=0$. Eqs. \eqref{deltau} and \eqref{deltax} implies that the integrated graviton momentum is orthogonal to the timelike vector $u_2$. Therefore, $q$ can never be on-shell and hit the pole $q^2 = 0$, allowing us to ignore the $i0^+$ prescription in the above three equations at this order.

\subsection{Stress-energy tensor}

As we explained in the previous section, we can compute the stress-energy tensor via a matching procedure, as depicted in eq.~\eqref{eq:match}. 
At leading order in $G$, particles move along straight trajectories, generating a static term. Using the Feynman rule written in eq. \eqref{eq:Fzero}, for body 1 we have
\begin{align}
\raisebox{18pt}{\Azerodiaglab} = \frac{m_1}{2 \mpl}  u_1^\rho u_1^\sigma \dd(k \cdot u_1) e^{i k \cdot b_1}\frac{P_{\rho\sigma\mu\nu}}{k^2}  \, .
\end{align}
Therefore, adding the symmetric contribution, we immediately find that
\be
\label{eq:Ts}
\tilde{T}^{\mu\nu}_{\rm LO}(k) = \sum_a m_a u_a^\mu u_a^\nu   e^{i k \cdot b_a} \dd (k \cdot u_a) \, .
\ee
The non-radiating nature of this contribution is manifest by the presence of the delta functions $\dd (k \cdot u_a)$.

At the next order, the stress-energy tensor $\tilde{T}^{\mu\nu}_{\rm NLO}$ is given by the sum of three contributions. The first is obtained when the worldline of the first body is deflected by the second one. Using the rule \eqref{eq:Fone}, we obtain 
\begin{align}
\label{eq:t1M}
\raisebox{18pt}{\Aonediaglab} & \!\!= \frac{m_1 }{2 \mpl}\int\!\! d \tau_1 e^{i k \cdot b_1+k \cdot u_1 \tau_1} \left( 2 \delta^{(1)} u_1^{(\rho}(\tau_1) u_1^{\sigma)} +i ( k \cdot  \delta^{(1)} x_1 (\tau_1) ) u_1^\rho u_1^\sigma \right) \frac{P_{\rho\sigma\mu\nu}}{k^2} \;.
\end{align}
The second contribution to $\tilde{T}^{\mu\nu}_{\rm NLO}$ is analogous to the first one, with the roles of the two bodies exchanged. The last contribution involves the cubic gravitational vertex and comes from evaluating the following diagram,
\begin{align}
\raisebox{30pt}{\Athreediaglab} & = -\frac{m_1 m_2}{4\mpl^3}\int_{q_1, q_2}\!\! \dd(q_1 \cdot u_1)\dd(q_2 \cdot u_2)\dd^{(4)}(q_1 + q_2 - k) \frac{e^{i q_1 \cdot b_1 + i q_2 \cdot b_2}}{q_1^2 q_2^2}\notag \\
\label{eq:Tmne}
&\qquad \times u_1^\alpha u_1^\beta  P_{\alpha \beta  \alpha_1 \beta_1} u_2^\rho u_2^\sigma P_{\rho \sigma  \alpha_2 \beta_2} V_3^{\alpha_1 \beta_1 \alpha_2 \beta_2 \alpha_3 \beta_3} \frac{P_{\alpha_3\beta_3\mu\nu}}{k^2} \, .
\end{align}
Note that this separation in three contributions is only for convenience and depends on the chosen gauge.

Summing everything together, we can thus write the next-to-leading-order stress-energy tensor  in Fourier space as 
\be
\label{eq:T}
\tilde{T}^{\mu\nu}_{\rm NLO}(k)  = \frac{m_1 m_2}{4 \mpl^2} \int_{q} \! \dd (q \cdot u_1) \dd (q \cdot u_2 - k \cdot u_2) \frac{e^{i q \cdot b}e^{i k \cdot b_2}}{q^2 (q-k)^2} \left[ {t}^{\mu\nu}_{\firstc} (q ,k) + {t}^{\mu\nu}_{\secondc}  (q ,k)+ {t}^{\mu\nu}_{\vdash} (q ,k) \right]\, , 
\ee
where ${t}^{\mu\nu}_{\firstc}$, ${t}^{\mu\nu}_{\secondc}$ and ${t}^{\mu\nu}_{\vdash}$ come respectively from eq. \eqref{eq:t1M}, its symmetric under $1 \leftrightarrow 2$ exchange and \eqref{eq:Tmne}. They are explicitly  given by
\begin{align}
\label{eq:t1}
t_{\firstc}^{\mu\nu}  (q,k) & \equiv q^2 \bigg\{ \frac{2  \gamma^2 -1 }{k \cdot u_1} (k - q)^{(\mu} u_1^{\nu)} - 4 \gamma u_1^{(\mu} u_2^{\nu)} + \left[ \frac{2\gamma^2 - 1}{2} \frac{k \cdot q}{(k \cdot u_1)^2} + 2\gamma \frac{k \cdot u_2}{ k \cdot u_1} + 1 \right] u_1^{\mu} u_1^{\nu} \bigg\}  \, , \\
\label{eq:t2}
t_{\secondc}^{\mu\nu}  (q,k) & \equiv (k - q)^2 \bigg\{ \frac{2  \gamma^2 - 1 }{k \cdot u_2} q^{(\mu} u_2^{\nu)} - 4 \gamma u_2^{(\mu} u_1^{\nu)} - \left[ \frac{2\gamma^2 - 1}{2} \frac{k \cdot q}{(k \cdot u_2)^2} - 2\gamma \frac{k \cdot u_1}{ k \cdot u_2} - 1 \right] u_2^{\mu} u_2^{\nu} \bigg\}  \, , 
\end{align}
and
\begin{align}
\label{eq:tc}
t_ {\vdash}^{\mu\nu} (q,k)& \equiv \frac{2\gamma^2 - 1}{2} \left[ k^\mu k^\nu  - 2 k^{(\mu} q^{\nu)} + q^\mu q^\nu\right] \!+\! \Big[ 2 ( k \cdot u_2)^2 - q^2 \Big] u_1^\mu u_1^\nu + 4 \gamma ( k \cdot u_2) (k - q)^{(\mu}u_1^{\nu)} \notag \\
& + \Big[ 2 ( k \cdot u_1)^2 -  (k - q)^2 \Big] u_2^\mu u_2^\nu - \eta^{\mu\nu} \left[ 2 \gamma (k \cdot u_1) (k \cdot u_2) +\frac{2\gamma^2 - 1}{4} \left((k-q)^2 +q^2 \right) \right] \notag \\
 &  + 4 \gamma ( k \cdot u_1) q^{(\mu}u_2^{\nu)}   + 2 \left[ \gamma \left((k-q)^2 +q^2 \right) - 2 (k \cdot u_1) (k \cdot u_2)  \right] u_1^{(\mu} u_2^{\nu)}
 \end{align}
The integral in eq.~\eqref{eq:T} is over the momentum of the graviton exchanged by the two  bodies. The two delta functions arise from the fact that we are taking the two bodies as non-propagating external sources. Note that similar integrals and delta functions appear when taking the classical limit of quantum observables in the scattering process between two massive particles. In this case, the integration variable $q$ is the difference between the momentum within the wavefunction and that in its conjugate (the so called momentum mismatch \cite{Kosower:2018adc}) while the delta functions arise from the on-shell constraints on the momenta of the scattering particles.
Also note that we have simplified the expression of the stress-energy tensor (particularly that in \eqref{eq:tc})
 by using momentum conservation as well as on-shell and harmonic gauge conditions, i.e.~$k^2 = 0$ and $k^\mu \tilde{h}_{\mu\nu}(k) = (1/2) k_\nu \eta^{\alpha\beta}\tilde{h}_{\alpha\beta}(k)$. This simplification 
 implies that the total stress-energy tensor in eq.~\eqref{eq:T} is transverse only for on-shell momenta, i.e.~$k_\mu \tilde{T}^{\mu \nu} \propto k^2 =0$ \cite{Mougiakakos:2021ckm}. 
 In App.~\ref{App:Tmunu} we report the complete expression of $\tilde{T}^{\mu\nu}_{\rm NLO}$ that is transverse also for off-shell momenta. Either expressions can be used to compute the emitted four-momentum, obviously leading to the same result.

Finally, we stress that  we have left implicit all the $i 0^+$ prescriptions in the denominators appearing either from the graviton propagator to specify the contour of integration in the complex $k^0$ plane or from the corrections to the straight motion of the two bodies, eqs.~\eqref{deltau} and \eqref{deltax}. 
Concerning the gravitons, in order to take into account only outgoing radiation one should impose retarded boundary conditions, e.g.~$\left[(q^0 + i0^+)^2 - |\vq|^2 \right]^{-1}$.(\footnote{For instance, radiation poles  play a key role for hereditary effects at higher orders \cite{Goldberger:2009qd}.}) In our case, however, these prescriptions are irrelevant because, as displayed by eq.~\eqref{eq:T}, the two delta functions ensure that the momenta $q$ and $q-k$ are orthogonal to one of the two four-velocities. Thus, these two momenta can  hit the pole only  in the trivial case  $q = q - k  = 0$. The same holds true for the matter poles $[k \cdot u_a \pm i 0^+]^{-1}$: since $k$ is on-shell  $k \cdot u_a$ can never vanish.

\subsection{Amplitude and waveform}\label{sec:ampl}

We can now compute the classical amplitude ${\cal A}_\lambda$ perturbatively in $G$ using eq. \eqref{amplitude}. The leading-order contribution is obtained from the static contribution of the stress-energy tensor, eq.~\eqref{eq:Ts},
\be
\label{waveformLO}
{\cal A}_\lambda^{\text{LO}} (k)  = - \sum_a \frac{m_a}{2 \mpl}  \epsilon^{\lambda *}_{\mu\nu} u_a^\mu u_a^\nu e^{i k \cdot b_a} \dd (k \cdot u_a) \;.
\ee

The next-to-leading-order amplitude is of order $G^{3/2}$. Analogously to what we did for the stress-energy tensor in eq.~\eqref{eq:T}, we can separate it  in three contributions
\be
\label{A2}
{\cal A}^{\text{NLO}}_\lambda (k)  = -\frac{m_1 m_2}{8 \mpl^3} \left({\cal A}_\lambda^{\firstc} (k) + {\cal A}_\lambda^{\secondc} (k) + {\cal A}_\lambda^{\vdash} (k) \right) \, ,
\ee
where the labels refer to the contribution with the same name given in eqs. \eqref{eq:t1}, \eqref{eq:t2} and \eqref{eq:tc}. Introducing the following set of integrals
\begin{align}
\label{eq:MI}
I^{\mu_1\dots\mu_n}_{(n)} & \equiv \int_q \dd(q\cdot u_1 - k\cdot u_1)\dd(q\cdot u_2)\frac{e^{- i q \cdot b}}{q^2}q^{\mu_1} \dots q^{\mu_n} \, , \\
\label{eq:MJ}
J^{\mu_1\dots\mu_n}_{(n)} & \equiv \int_q \dd(q\cdot u_1 - k\cdot u_1)\dd(q\cdot u_2)\frac{e^{- i q \cdot b}}{q^2(k - q)^2}q^{\mu_1} \dots q^{\mu_n} \, ,
\end{align}
we have explicitly that
\begin{align}
\label{eq:Al1_MI}
{\cal A}_\lambda^{\firstc} (k) & =\! \epsilon^{\lambda *}_{\mu\nu}\bigg\{\!\frac{2 \gamma^2 - 1}{k\cdot u_1} I_{(1)}^\mu u_1^\nu - 4I_{(0)}\gamma u_1^\mu u_2^\nu \notag \\
& \hspace{3cm} - \!\left[\frac{2\gamma^2 - 1}{2}\frac{k \cdot I_{(1)}}{(k \cdot u_1)^2} - \left(2\gamma\frac{k\cdot u_2}{k\cdot u_1}+1\right)\! I_{(0)}\right]\! u_1^\mu u_1^\nu \bigg\}e^{i k \cdot b_1} \, ,\\
\label{eq:Al2_MI}
{\cal A}_\lambda^{\secondc} (k)  & = \left. {\cal A}_\lambda^{\firstc} (k)\right|_{1\leftrightarrow 2} \, , \\ 
\label{eq:Ac_MI}
{\cal A}_\lambda^{\vdash} (k) & = \epsilon^{\lambda *}_{\mu\nu}\bigg\{ \frac{2\gamma^2 - 1}{2} J_{(2)}^{\mu\nu} +\left(2 (k \cdot u_2)^2 J_{(0)}-I_{(0)}\right)u_1^\mu u_1^\nu + 4 \gamma k \cdot u_2 J_{(1)}^\mu u_1^\nu \notag \\
&\qquad\qquad\qquad  -\eta^{\mu\nu}\left[\gamma (k \cdot u_1)(k \cdot u_2)J_{(0)} +\frac{2\gamma^2 - 1}{4}I_{(0)}\right]  \notag \\
& \qquad\qquad\qquad  + 2\left[\gamma I_{(0)} -  (k \cdot u_1)(k \cdot u_2)J_{(0)}\right]u_1^\mu u_2^\nu  \bigg\}e^{i k \cdot b_1} +(1\leftrightarrow 2) \, .
\end{align}
We stress that in eqs \eqref{eq:Al2_MI} and \eqref{eq:Ac_MI} one needs to exchange the label $1 \leftrightarrow 2$ also inside the definition of the integrals $I^{\mu_1\dots\mu_n}_{(n)}$ and $J^{\mu_1\dots\mu_n}_{(n)}$. At this point, we are left with solving the integrals in eqs.~\eqref{eq:MI} and \eqref{eq:MJ}. As discussed in \cite{Mougiakakos:2021ckm}, for the set $I^{\mu_1\dots\mu_n}_{(n)}$ it is possible to find an analytic solution in terms on known function while for the set $J^{\mu_1\dots\mu_n}_{(n)}$ the best we can do is to write them as one dimensional integrals over a Feynman parameter.
See App.~\ref{app:coeff} for  details of the calculations.

The next-to-leading-order amplitude takes a rather compact form if we consider the polarization tensor in the transverse-traceless (TT) gauge, i.e.,
\begin{align}
\label{eq:TTgauge}
\epsilon^\lambda_{0\mu} = 0 \, , & & k^\nu \epsilon^\lambda_{\mu\nu} = 0 \, , & & \epsilon^\lambda_{\mu\nu}\eta^{\mu\nu} = 0 \, ,
\end{align}
and we choose the reference frame in which one of the two bodies, say body 2, is at rest, i.e.
\begin{align}
u_2^\mu = \delta^{\mu}_{0} \, , & & u_1^\mu = \gamma v^\mu = (\gamma , \sqrt{\gamma^2 -1}\, \ve_v) \, , & & b_2^\mu = 0 \, , & & b_1^\mu = b^\mu = (0, |\vb| \ve_b ) \, ,
\label{eq:frame}
\end{align}
where $\ve_v$ and $\ve_b$ are mutually orthogonal unitary vectors, $\ve_v \cdot \ve_b = 0$, $|\ve_v| = |\ve_b| = 1$. 
With this choice ${\cal A}_\lambda^{\secondc} (k) = 0$ and all but one term in the symmetric contribution in eq.~\eqref{eq:Ac_MI} drop. Finally, parametrizing the graviton four-momentum as $k^\mu = \omega n^\mu$, with $n^\mu$ given in eq. \eqref{eq:defn}, and defining
\begin{align}
z \equiv \frac{\gamma |\vb| \omega}{\sqrt{\gamma^2 -1}}  \, , \qquad\qquad f(y) \equiv  \sqrt{ (1-y)^2 (n \cdot v)^2 + 2 y    (1-y) (n \cdot v) + y^2/\gamma^2   } \; ,
\end{align}
one can write the next-to-leading-order amplitude in a compact form as 
\begin{align}
 {\cal A}_\lambda^{\text{NLO}} (k)  &= - \frac{ G m_1 m_2}{ \mpl \sqrt{\gamma^2 - 1} }  \epsilon^{*\lambda}_{ij}    \ve_I^i \ve_J^j    { A}_{IJ} (k) e^{i k \cdot b} \; ,
 \label{eq:NLO_A}
\end{align}
where $I, J = {v, b}$ and the coefficients $A_{IJ}$ are explicitly given by\footnote{We thank Paolo Di Vecchia, Carlo Heissenberg, Rodolfo Russo and Gabriele Veneziano for 
correspondence and comparison of the waveform.} \cite{Mougiakakos:2021ckm}
\begin{align}
{ A}_{vv} & =   c_{1} K_0 \big( z ( n \cdot v) \big) +  i c_2  \Big[ K_1 \big( z ( n \cdot v) \big)    - i \pi \delta \big( z ( n \cdot v) \big)  \Big]  \notag \\
  & \quad +  \int_0^1  \! dy \, e^{i y \vk \cdot \vb }  \Big[     d_{1} (y) z  K_1 \big(  z f(y) \big)  + c_0 K_0 \big( z f(y) \big) \Big]   \;, \label{eqs:A1} \\
{ A}_{vb} &  =   i c_{0}  \Big[ K_1 \big( z ( n \cdot v ) \big)  - i   \pi \delta \big( z ( n \cdot v) \big)  \Big]  +   i \int_0^1 \! dy \, e^{i y \vk \cdot \vb } d_{2} (y)  z  K_0 \big( z f(y) \big)   \; , \label{eqs:A2} \\
{ A}_{bb}  & =   \int_0^1 \! dy \, e^{i y \vk \cdot \vb }      d_{0} (y) z K_1 \big( z f(y) \big)  \; , \label{eqs:A3}
\end{align}
where the coefficients $c$ and $d$ are given by
\be
\begin{split}
c_0  = & \ 1- 2 \gamma^2 \;, \qquad c_{1} =  - c_0 +\frac{3 - 2 \gamma^2}{n \cdot v} \;, \qquad c_2  =  \frac{\sqrt{\gamma^2 -1 }}{\gamma} c_{0}    \frac{  \vn \cdot \ve_b}{  n \cdot v} \;, \\
d_{0} (y) & = f(y) c_0  \;, \\
d_{1} (y) & =   \frac{\gamma^2 -1}{\gamma^2}\frac{ 4 \gamma^2 (y-1) (n \cdot v) -  c_0 (y-1)^2  -2 y -1 }{f(y)}  - d_0 (y)\;, \\
d_{2} (y) & =  - 1 + (1-y)c_0 (n \cdot v -1)   \;.
\end{split}
\ee

In Ref.~\cite{Mougiakakos:2021ckm} we have compared this amplitude with previous results \cite{Kovacs:1978eu}, even if only in particular limits. But eq.~\eqref{eq:NLO_A} can be integrated in the frequency using \eqref{waveform} and this results in the next-to-leading-order waveform in direct space, which fully agrees with that derived long ago by Kovacs and Thorne \cite{Kovacs:1978eu} (see also \cite{Jakobsen:2021smu}). The emission amplitude can  be also used to compute radiated observables, such as the linear and angular momentum loss by the system. 
The emitted  angular momentum starts at order ${\cal O}(G^2)$ and involves  the leading-order amplitude \eqref{waveformLO} and only the soft limit of the next-to-leading order amplitude. In the soft limit, the integrals in eqs.~\eqref{eqs:A1}--\eqref{eqs:A3} can be performed and the waveform can be given analytically and used to compute the angular momentum  \cite{Mougiakakos:2021ckm}, which agrees with previous results \cite{Damour:2020tta}. However, the leading-order radiated four-momentum involves the full next-to-leading order amplitude. 
Due to the involved structure of the integrals over the Feynman parameter $y$, the waveform can only be analytically computed expanding \eqref{eq:NLO_A} for small velocities. In \cite{Mougiakakos:2021ckm} we have used this expansion to derive the emitted four-momentum up to ${\cal O}(v^8)$. 
In the next section, we will take another route which dispenses with the need for an analytical expression of ${\cal A}^{\text{NLO}}_\lambda$
and which leads directly to the full emitted momentum.

\section{Radiated 4-momentum as 2-loop integral} \label{Sec:2loop_int}
\label{Sec3}


In term of the classical amplitude of graviton emission ${\cal A}_\lambda (k)$, the radiated total momentum $P^\mu_{\text{rad}}$ is given by \cite{Goldberger:2017vcg,Mougiakakos:2021ckm}
\be
\label{Pmu}
P^\mu_{\text{rad}}  = \sum_\lambda \int_k   \ddp (k^2) k^\mu \left| {\cal A}_\lambda (k) \right|^2 \;,
\ee
where $ \frac{d^4 k}{(2 \pi)^4} \ddp (k^2) $ is the Lorentz-invariant graviton on-shell phase-space measure. 
The differential probability  of emission of one graviton with polarization $\lambda$
is
\be 
\label{numberg}
d N_\lambda = \frac{d^{3} k}{(2 \pi)^{3} } \int  \frac{d k^0 }{2 \pi}   \ddp(k^2) \left| {\cal A}_\lambda (k) \right|^2  \;.
\ee
This quantity is not well-defined classically: if we interpret  $\vk$ and $k^0$ in these expressions as classical wave-vector and frequency, respectively,  and we restore $\hbar \neq 1$, we must add a factor of $\hbar^{-1}$ in front of the right-hand side,\footnote{Restoring $\hbar \neq 1$, the amplitude is defined as $i {\cal A}_\lambda( k) = -  i \sqrt{8 \pi G} \epsilon^{ * \lambda}_{\mu \nu}  \tilde T^{\mu \nu}(k)$. Distinguishing units of energy and length, denoted respectively by $[M]$ and $[L]$, it has units  $[M]^{1/2} [L]^{3/2}$. The factor  $\hbar^{-1}$ in eq.~\eqref{numberg} restores the correct dimensions of the right-hand side, making it dimensionless.} which  shows that the number of emitted gravitons is divergent in the classical limit $\hbar \to 0$.
However, inserting the four-momentum of the graviton  $k^\mu$  gives a finite quantity in the classical limit and integrating over all gravitons  we obtain the total classical emitted momentum above.

Pictorially, eq.~\eqref{Pmu} can be represented by
\be
\label{Pmu2}
P^\mu_{\text{rad}}  = \sum_\lambda \int_k   \ddp(k^2) k^\mu \left| \raisebox{2pt}{\scalebox{0.9}{\Asqdiaglab}} \right|^2 \;,
\ee
where the on-shell  amplitude on the right-hand side is non perturbative in $G$.
Here we focus on the leading-order emitted momentum and therefore we expand the amplitude in powers of $G$ as 
\be
\raisebox{2pt}{\scalebox{0.9}{\Asqdiag}} =  \raisebox{24pt}{\scalebox{0.9}{\Azerodiag}}  \ + \!\! \raisebox{27pt}{\scalebox{0.9}{\Azerotwodiag}} \ + \!\! \raisebox{24pt}{\scalebox{0.9}{\Aonediag}} \ + \!\! \raisebox{24pt}{\scalebox{0.9}{\Atwodiag}} \ + \!\! \raisebox{24pt}{\scalebox{0.9}{\Athreediag}}  + \cdots
\ee
The first two diagrams on the right-hand side, of order $G^{1/2}$, are static (they are proportional to $\dd( k \cdot u_a)$) and when multiplied by $k^\mu$ they do not contribute to the emitted power. Therefore, the leading order contribution to the radiated power comes from squaring the last three diagrams, of order $G^{3/2}$,
\be 
\label{Pmu3}
k^\mu \left| \raisebox{2pt}{\scalebox{0.9}{\Asqdiaglab}}  \right|_{\rm LO}^2     =   k^\mu  \left| \raisebox{24pt}{\scalebox{0.9}{\Aonediagl}}  \ +  \ \raisebox{24pt}{\scalebox{0.9}{\Atwodiagl}}  \ +\   \raisebox{24pt}{\scalebox{0.9}{\Athreediagl}}      \right|^2 \;.
\ee

As explained in the previous section, we were  unable to solve the integral  in eq.~\eqref{eq:T} in the momentum of the graviton exchanged between the particles, $q$, and express the full amplitude (and waveform) in terms of known functions. 
In the following we adopt a different strategy to compute the right-hand side of eq.~\eqref{Pmu}. 
Another pictorial interpretation of this equation is
\be
\label{Pmu4}
P^\mu_{\text{rad}}  = \sum_\lambda \int_k   k^\mu  \;  \raisebox{2pt}{\scalebox{0.8}{\Vtovdiag}}    \;,
\ee
where we interpreted $\ddp(k^2)$ as a cut propagator so that the modulo squared of the amplitude has been replaced by a vacuum-to-vacuum diagram with a cut. We also depict explicitly the flow of the momentum $k^\mu$ as dictated by the positive energy theta function in $\ddp(k^2)$. 
From \eqref{Pmu3},  at leading order we expect four different  cut topologies on the right-hand side, coming from the different ways of combining the three contributions in the modulo squared of the amplitude at leading order, denoted here by M, N, IY and H type, i.e.,
\be
\label{eq:2loop}
\left( \raisebox{2pt}{\scalebox{0.8}{\Vtovdiag}} \right)_{\rm LO} = \underset{\text{M}}{\raisebox{24pt}{\scalebox{0.8}{\Mdiag}}}   \ +  \  \underset{\text{N}}{\raisebox{24pt}{\scalebox{0.8}{\Ndiag}} }  \  + \   \underset{\text{IY}}{\raisebox{24pt}{\scalebox{0.8}{\IYdiag}}} \ + \   \underset{\text{H}}{\raisebox{24pt}{\scalebox{0.8}{\Hdiag}}}  \ + (1 \leftrightarrow 2) \;.
\ee 
Then, following a technique  employed in high-energy physics computations known under the name of {\em reverse unitarity} \cite{Anastasiou:2002yz,Anastasiou:2002qz,Anastasiou:2003yy,Anastasiou:2015yha}, we can differentiate the cut propagator of eq.~\eqref{Pmu4}  like normal virtual propagators and apply the standard procedure of IBP reduction  \cite{Tkachov:1981wb,Chetyrkin:1981qh, Smirnov:2012gma} to the resulting integral.

In practice, we rewrite the modulo squared of the amplitude as
\be
\label{AtoT}
\sum_\lambda | {\cal A}^{\text{NLO}}_\lambda (k) |^2  =  \frac{1}{4 \mpl^2} P_{\alpha \beta \rho \sigma} \tilde T_{\rm NLO}^{\alpha \beta}(k) \tilde T_{\rm NLO}^*{}^{\rho \sigma}(k) \;,
\ee
where we have used eq.~\eqref{amplitude} and
\be
\label{eq:sum_pol}
\sum_\lambda \epsilon^{\lambda *}_{\alpha \beta} \epsilon^{\lambda}_{\rho \sigma} = \eta_{\alpha (\beta} \eta_{\sigma) \rho} - \frac{1}{ 2} \eta_{\alpha \beta}\eta_{\rho\sigma} \equiv P_{\alpha \beta\rho\sigma} \, .
\ee
Expanding the right-hand side of eq.~\eqref{AtoT} using \eqref{eq:T}, we obtain
\be
\begin{split}
\label{AN}
\sum_\lambda | {\cal A}^{\text{NLO}}_\lambda (k) |^2  = \frac{m_1^2 m_2^2}{64 \mpl^{6}} \int_{q_1,q_2}\!\!\!\dd (q_1 \cdot u_1) \dd (q_1 \cdot u_2 - k \cdot u_2) & \dd (q_2 \cdot u_1) \dd (q_2 \cdot u_2 - k \cdot u_2)   \\
& \times \frac{e^{i (q_1 - q_2) \cdot b} {\cal N }(q_1,q_2,k) }{q_1^2 q_2^2 (k - q_1)^2 (k - q_2)^2} \;,
\end{split}
\ee
where the numerator ${\cal N}$ can be organized  in terms of the contributions from the four topologies above and is explicitly defined as
\begin{equation}
 {\cal N}(q_1,q_2,k) \equiv \Big(t_{\firstc}^{\mu\nu}(q_1,k) +  t_{\secondc}^{\mu\nu}(q_1,k) +  t_{\vdash}^{\mu\nu}(q_1,k)\Big) P_{\mu\nu\rho\sigma}\Big(t_{\firstc}^{\rho\sigma}(q_2,k) +  t_{\secondc}^{\rho\sigma}(q_2,k) +  t_{\vdash}^{\rho\sigma}(q_2,k)\Big)^* \, .
\end{equation}
Finally, replacing the modulo squared of the amplitude in eq.~\eqref{Pmu} using \eqref{AN} and renaming
\begin{align}
q_1^\mu = \ell_1^\mu  \;, & &  q_2^\mu = \ell_1^\mu - q^\mu \, , & &  k^\mu = \ell_1^\mu + \ell_2^\mu - q^\mu \, ,
\end{align}
we obtain
\be
\begin{split}
\label{eq:prad_2loop}
P^\mu_{\text{rad}}  = \frac{m_1^2 m_2^2}{64 \mpl^{6}} \int_q \dd (q\cdot u_1) \dd (q\cdot u_2) e^{i q \cdot b} &  \int_{\ell_1, \ell_2} \dd_+ ((\ell_1+\ell_2 - q)^2)  \dd (\ell_1\cdot u_1)\dd (\ell_2\cdot u_2) \\ & \times \frac{\big(\ell_1^\mu + \ell_2^\mu - q^\mu \big) {\cal N} (\ell_1, \ell_2, q) }{\ell_1^2 \ell_2^2 (\ell_1 - q)^2 (\ell_2 - q)^2 } \, .
\end{split}
\ee
At this stage, we have  rewritten the total four-momentum emitted as a cut 2-loop integral, followed by a Fourier transform from $q$ to $b$-space. The advantage of this procedure is that we can now solve the 2-loop integral  all at once, making use of the powerful   computational tools routinely employed in high-energy physics---IBP reduction into master integrals  \cite{Tkachov:1981wb,Chetyrkin:1981qh, Smirnov:2012gma} and differential equation methods \cite{Kotikov:1990kg,Bern:1992em,Gehrmann:1999as,Henn:2013pwa} to solve the latter---without the need of deriving the Fourier-space gravitational waveform. This is analogous to the calculations recently performed in \cite{Herrmann:2021lqe,Herrmann:2021tct,DiVecchia:2021bdo}. However, here   
we do not have to consider any intermediate quantum or super-classical contributions in our integrals: the amplitude ${\cal A}_\lambda$ is a classical observable from the start.   
Before computing the contribution from each of the topologies in eq.~\eqref{eq:2loop} we need to discuss the master integrals that we will need to solve the associated two-loop integrals. This is what we turn to now.

\section{Solving the integral }
\label{Sec4}

\subsection{Master integrals}

As explained in \cite{Mougiakakos:2021ckm}, since the modulo squared of the amplitude is symmetric under the exchanged $\vk\cdot \vb \rightarrow -\vk\cdot \vb$, the four-momentum cannot depend on the spatial direction $b^\mu$. Moreover, the energy measured in the frame of one body is the same as the one measured in the  frame of the other one, hence  the final result must be proportional to $u_1^\mu + u_2^\mu$. 
We can write this result as 
\cite{Herrmann:2021lqe}
\be 
\label{totalmom}
P^{\mu}_{\rm rad} = \frac{G^3 m_1^2 m_2^2}{|\vb|^3} \frac{u_1^\mu + u_2^\mu}{\gamma + 1}  {\cal E}(\gamma)+ {\cal O}(G^4) \; .
\ee

We focus on the computation of ${\cal E} (\gamma)$, which we can be extracted by multiplying both eqs.~\eqref{eq:prad_2loop} and \eqref{totalmom} by $u_1^\mu + u_2^\mu$ and comparing their right-hand sides. Therefore, we get
\be 
\label{eq:Arad}
{\cal E} (\gamma) =  512 \pi^3 |\vb|^3 \int_q \dd (q\cdot u_1) \dd (q\cdot u_2) e^{i q \cdot b} {\sqrt{-q^2}}  {\cal I} (\gamma) \, ,
\ee
with
\be
\begin{split}
\label{eq:I}
{\cal I} (\gamma) \equiv \frac{1}{2\sqrt{-q^2}} \int_{\ell_1, \ell_2} & \ddp ((\ell_1+\ell_2 - q)^2) \dd (\ell_1\cdot u_1)\dd (\ell_2\cdot u_2) \\
& \times \frac{\big( \ell_1 \cdot u_2 + \ell_2 \cdot u_1 \big) {\cal N} (\ell_1, \ell_2, q) }{\ell_1^2 \ell_2^2 (\ell_1 - q)^2 (\ell_2 - q)^2 } \;.
\end{split}
\ee
Notice that both ${\cal E}$ and ${\cal I}$ are dimensionless and  only dependent on $\gamma= u_1 \cdot u_2$. 
Indeed,
the  2-loop integral on the right-hand side of eq.~\eqref{eq:I} has dimension one. It can only depend on $q^2$ and $\gamma= u_1 \cdot u_2$ because $q\cdot u_1=q\cdot u_2=0$ by the delta functions in eq.~\eqref{eq:Arad} and no poles are expected.  Since  only $q$ is dimensionful, it must scale as $\sqrt{-q^2}$, which is compensated by the  prefactor. 
The integral in eq.~\eqref{eq:Arad} has dimension three and since the  only dimensionful parameter is  $b$, it must scale like $ |\vb|^{-3}$. This removes the $ |\vb|$-dependence on   the right-hand side making ${\cal E} $  dimensionless.

We now discuss how to  simplify and solve the 2-loop integral ${\cal I}$. Use the notation \cite{Parra-Martinez:2020dzs, DiVecchia:2021bdo, Herrmann:2021tct}
\be
\label{eq:basis1}
 \rho_1  = 2 \ell_1 \cdot u_1\, ,  \qquad \rho_2 = - 2 \ell_1 \cdot u_2\, , \qquad \rho_3 = - 2 \ell_2 \cdot u_1\, , \qquad \rho_4 = 2 \ell_2 \cdot u_2\, ,
\ee
and
\be
\label{eq:basis2}
\rho_5  = \ell_1^2 \, , \qquad \rho_6 = \ell_2^2 \, , \qquad \rho_7 = (\ell_1 + \ell_2 - q)^2 \, , \qquad \rho_8 = (\ell_1 - q)^2 \, , \qquad \rho_9 = (\ell_2 - q)^2 \, , 
\ee
and rewrite it as
\be
\label{eq:Arad3}
{\cal I} (\gamma) =  -\frac{1}{\sqrt{-q^2}} \int_{\ell_1, \ell_2} \ddp (\rho_7) \dd (\rho_1) \dd (\rho_4)\frac{\big(\rho_2 + \rho_3\big) {\cal N} (\rho_1, \ldots, \rho_9)}{\rho_5 \rho_6   \rho_8 \rho_9} \, .
\ee
We use dimensional regularization and extend the 4-dimensional integration  to $d$ spacetime dimensions, i.e.
\begin{align}
\int_{\ell_1, \ell_2} \equiv \int \frac{d^d \ell _1}{(2\pi)^d}\frac{d^d \ell _2}{(2\pi)^d} \, , \qquad d = 4 - 2 \varepsilon \;.
\end{align} 
Moreover,  reverse unitarity  \cite{Anastasiou:2002yz,Anastasiou:2002qz,Anastasiou:2003yy,Anastasiou:2015yha} allows to treat the three delta functions involving $\rho_1$, $\rho_4$ and $\rho_7$    as cut propagators and apply IBP techniques fixing the boundary conditions according to these cuts.

In particular, we formally replace the three delta functions by cut propagators and we underline them to distinguish them from the standard ones,
\be
\ddp (\rho_7) \to \frac1{\underline{\rho_7}} \;, \qquad  \dd (\rho_1) \to \frac1{\underline{\rho_1}} \;, \qquad  \dd (\rho_4) \to \frac1{\underline{\rho_4}} \;.
\ee 
Then, ${\cal I}$ is  given as a linear combination of integrals of the form   
\be 
G_{\underline{i_1}, i_2, i_3, \underline{i_4}, i_5, i_6, \underline{i_7}, i_8, i_9} = \int_{\ell_1, \ell_2} \frac{1}{\underline{\rho}_{1}^{i_1} \rho_{2}^{i_2} \rho_{3}^{i_3} \underline{\rho}_{4}^{i_4} \rho_{5}^{i_5} \rho_{6}^{i_6} \underline{\rho}_{7}^{i_7} \rho_{8}^{i_8} \rho_{9}^{i_9}} \, .
\ee
 With the help of \texttt{LiteRed} \cite{Lee:2012cn, Lee:2013mka}, a \texttt{Mathematica} package performing the IBP reduction to master integrals,
  this combination can be reduced to  the following four master \linebreak integrals:  $f_1 \equiv \sqrt{-q^2} G_{\underline{2}, 0, 0, \underline{1}, 0, 1, \underline{1}, 0, 1}$, $f_2 \equiv \sqrt{-q^2}  G_{\underline{2}, 0, 0, \underline{1}, 0, 0, \underline{1}, 1, 1}$, $f_3 \equiv \sqrt{-q^2}  G_{\underline{1}, 0, 1, \underline{1}, 0, 0, \underline{1}, 1, 1}$ and $f_4 \equiv (-q^2)^{5/2}  G_{\underline{2}, 0,  0, \underline{1}, 1, 1, \underline{1}, 1, 1}$. Once the master integrals are known, we can cast ${\cal I}$ as a linear combination of these.
Note that since we are considering a cut two-loop integration, one must use the \texttt{CutDS} option in \texttt{LiteRed} in order to perform the correct IBP reduction. The set of propagators in eqs.~\eqref{eq:basis1} and \eqref{eq:basis2} and the four master integrals above are enough to solve our four topologies in eq.~\eqref{eq:2loop}. 

At this point, we can use the differential equation methods \cite{Kotikov:1990kg,Bern:1992em,Gehrmann:1999as,Henn:2013pwa} to solve these integrals. It is convenient to replace the dependence on $\gamma$ of the master integrals by
that  on the  kinematic variable $x$, defined by 
$x \equiv \gamma- \sqrt{\gamma^2- 1}$ \cite{Parra-Martinez:2020dzs}. The following relations derived from this definition will be useful later,
\begin{align}
\label{eq:defx}
 \gamma = \frac{1+ x^2}{2 x} \, , \qquad \sqrt{\gamma^2 -1} = \frac{1- x^2}{2 x} \, .
\end{align} 
Differentiating with respect to $x$, one  realizes that the above integrals  satisfy a system of differential equations of the  form
\be
\partial_x \vec f(x,\varepsilon) = F(x, \varepsilon) \vec f(x,\varepsilon) \;,
\ee
where $\vec f   \equiv \{ f_1,f_2,f_3,f_4\}$   and $F(x, \varepsilon)$ is a matrix of rational coefficients. The properties of Feynman integrals ensure that  the above system has  only  regular singularities, i.e., it is a Fuchsian system of differential equations. To solve this equation, it is convenient to find a basis $\vec{g} = \{g_1, g_2, g_3, g_4\}$ such that 
the  differential equation is in canonical form \cite{Henn:2013pwa, Caron-Huot:2014lda,Henn:2014qga}, i.e., 
\be
\partial_x \vec g(x,\varepsilon) =\varepsilon A(x ) \vec g(x,\varepsilon) \;.
\ee
A system of this form can be trivially solved in terms of polylogarithms as a Laurent series in $\varepsilon$.
The transformation between the basis $\vec{f} $ and $\vec{g} $ can be obtained with the help of the package \texttt{Fuchsia} \cite{Gituliar:2016vfa, Gituliar:2017vzm},  implementing the Lee algorithm \cite{Lee:2014ioa}.

The canonical basis of master integrals reads 
\begin{align}
\label{eq:g1}
g_1 & =  \sqrt{-q^2} G_{\underline{2}, 0, 0, \underline{1}, 0, 1, \underline{1}, 0, 1} \, , \\
\label{eq:g2}
g_2 & = \sqrt{-q^2} G_{\underline{2}, 0, 0, \underline{1}, 0, 0, \underline{1}, 1, 1} \, , \\
\label{eq:g3}
g_3 & = \varepsilon  \sqrt{-q^2} \sqrt{\gamma^2 - 1} G_{\underline{1}, 0, 1, \underline{1}, 0, 0, \underline{1}, 1, 1} \, , \\
\label{eq:g4}
g_4 & = \big(\sqrt{-q^2}\big)^5 \frac{\gamma - 1}{8} G_{\underline{2}, 0,  0, \underline{1}, 1, 1, \underline{1}, 1, 1} + \sqrt{-q^2}\frac{1- 2\varepsilon (2 +3 \gamma)}{12 (1 + 2\varepsilon)} G_{\underline{2}, 0,  0, \underline{1}, 0, 0, \underline{1}, 1, 1} \notag \\
&\qquad\qquad\qquad\qquad\qquad + \frac{2\varepsilon}{(1+ 2\varepsilon) (1 + \gamma)} \sqrt{-q^2} G_{\underline{2}, 0, 0, \underline{1}, 0, 1, \underline{1}, 0, 1} \, , 
\end{align}
which satisfies the following canonically normalized differential equation,
\begin{align}
\frac{d}{d x} \vec{g} (x,\varepsilon) = \varepsilon
\begin{pmatrix}
- \frac{2(1 + x^2)}{x (x^2 -1)} & 0 & 0 & 0 \\
0 & \frac{2 (1 - 4x +x^2)}{x (x^2 -1)} & 0 & 0 \\
0 & \frac{1}{x} & 0 & 0 \\
-\frac{4}{x^2 -1} & \frac{7 + 10x +7x^2}{6x (x^2 -1)} & 0 & -\frac{4}{x^2 -1}
\end{pmatrix}
\vec{g} (x,\varepsilon) \, , \qquad \vec{g}  \equiv \begin{pmatrix}
g_1 \\ g_2 \\ g_3 \\ g_4
\end{pmatrix} \, .
\end{align}
This can be equivalently written as
\be
\label{eq:diffeq_log}
d \vec{g} = \varepsilon \big[A_0\,d\!\log x + A_{+1}\,d\!\log (x + 1) + A_{-1}\,d\!\log (x - 1)\big] \;,
\ee
with
\begin{align}
A_0 & = \begin{pmatrix}
-2 & 0 & 0 & 0 \\
0 & -2 & 0 & 0 \\
0 & 1 & 0 & 0 \\
0 & \frac{7}{6} & 0 & 0 \\
\end{pmatrix} \, , \qquad \quad
A_{+1} = \begin{pmatrix}
2 & 0 & 0 & 0 \\
0 & 6 & 0 & 0 \\
0 & 0 & 0 & 0 \\
2 & \frac{1}{3} & 0 & 2 \\
\end{pmatrix} \, ,
\qquad \quad
A_{-1} = \begin{pmatrix}
2 & 0 & 0 & 0 \\
0 & -2 & 0 & 0 \\
0 & 0 & 0 & 0 \\
-2 & 2 & 0 & -2 \\
\end{pmatrix} \, .
\end{align}
This differential equation can be solved perturbatively in $\varepsilon$ \cite{DiVecchia:2021bdo}, i.e., for each $j = 1, \dots, 4$,
\be 
g_j = \frac{1}{(- q^2)^{2\varepsilon}} \sum_k g_j^{(k)} \varepsilon^k \, .
\ee

The last ingredient  are the boundary conditions of the differential equation \eqref{eq:diffeq_log}. These can be found by solving the master integrals in the near static limit, i.e.~for $\gamma \to 1$ (or $x \to 1$). We give an explicit derivation of these boundary conditions in App.~\ref{App:BC} and we report here the results:
\be
\label{eq:BC}
g_1 |_{\gamma \to 1} =   g_2 |_{\gamma \to 1}= 12 g_4|_{\gamma \to 1}  = -\frac{C_\text{BC}}{(4\pi)^{4-2\varepsilon}}   \, , \qquad  g_3|_{\gamma \to 1} = 0 \, ,  
\ee
where
\be
\label{eq:BC2}
C_{\text{BC}} = \sin( \pi \varepsilon)\left(\frac{1}{(-q^2)(1 - x)}\right)^{2\varepsilon} \frac{\sqrt{\pi}\Gamma(1+\varepsilon)\Gamma(1-\varepsilon)\Gamma(\frac{1}{2}+2\varepsilon)\Gamma(\frac{1}{2}-2\varepsilon)^{2}}{\varepsilon \Gamma(1-4\varepsilon)} \, .
\ee
For the following computation we just need the solutions up to order $\varepsilon$. These are explicitly given by
\begin{align}
\label{eq:sol0}
g_1^{(0)} & = -\frac{1}{256 \pi} \, , \qquad\quad g_2^{(0)} = -\frac{1}{256 \pi} \, , \qquad\quad g_3^{(0)} = 0 \, , \qquad\quad g_4^{(0)} = -\frac{1}{3072 \pi} \, , \\
\label{eq:sol1_1}
g_1^{(1)} & = \frac{\gamma_\text{E} - \log(4\pi)}{128 \pi} +\frac{1}{128 \pi}\left[ \log\left(\frac{1-x}{4}\right) - \log(x) + \log\left(\frac{1+ x}{2}\right) \right] \, , \\
\label{eq:sol2_1}
g_2^{(1)} & = \frac{\gamma_\text{E} - \log(4\pi)}{128 \pi} +\frac{1}{128 \pi}\left[ \log\left(\frac{1-x}{4}\right) + \log(x) -3 \log\left(\frac{1+ x}{2}\right) \right] \, , \\
\label{eq:sol3_1}
g_3^{(1)} & = -\frac{1}{256 \pi} \log(x)\, , \\
\label{eq:sol4_1}
g_4^{(1)} & = \frac{\gamma_\text{E} - \log(4\pi)}{1536 \pi} +\frac{1}{1536 \pi}\left[ \log\left(\frac{1-x}{4}\right) + 7 \log(x) -15 \log\left(\frac{1+ x}{2}\right) \right] \, ,
\end{align}
where $\gamma_{\rm E}$ is the Euler-Mascheroni constant.
Up to a different normalization of the loop integrals,\footnote{In the QCD/amplitude  literature it is common practice to remove a factor of
$i (4 \pi)^{\varepsilon - 2} e^{-\varepsilon \gamma_{\rm E}}$ from the normalization of the integrals. Here we do not use this convention.} these agree with  \cite{DiVecchia:2021bdo}. 
To conclude this section, we stress that before solving the master integrals $\vec g$ in the static limit $x \to 1$ one must recast the cut propagators as delta functions. In practice, the resulting integrals are hard to solve. It is more convenient to relate these cut  integrals to their non-cut versions using the so-called Cutkosky's rules \cite{Cutkosky:1960sp}. The latter can then be solve using standard loop integral techniques.  We explain all this in details in App~\ref{sec:cut}.

\subsection{Computing the four topologies}

We have now all the ingredients to compute the leading-order  radiated momentum $P_{\rm rad}^{\mu}$. As mentioned in the previous section, we will focus on computing ${\cal E} (\gamma)$ defined eq.~\eqref{eq:Arad},  splitting the computation in four contributions coming from the four topologies in  eq.~\eqref{eq:2loop}, i.e.,
\be 
\label{eq:Arad_2}
{\cal E} = {\cal E}_{\text{M}} + {\cal E}_{\text{N}} + {\cal E}_{\text{IY}} + {\cal E}_{\text{H}} + (u_1 \leftrightarrow u_2)  \; ,
\ee
where
\be
\label{EI}
{\cal E}_{I}(\gamma)  \equiv 512 \pi^3 |\vb|^3 \int_q \dd (q\cdot u_1) \dd (q\cdot u_2) e^{i q \cdot b}  \sqrt{-q^2} {\cal I}_I (\gamma) \;, 
\ee
and
\be
{\cal I}_I (\gamma)  \equiv -\frac{1}{\sqrt{-q^2}}\int_{\ell_1, \ell_2} \frac{\big(\rho_2 + \rho_3\big) {\cal N}_{I} (\rho_1, \ldots, \rho_9)}{\underline{\rho}_1 \underline{\rho}_4 \rho_5 \rho_6 \underline{\rho}_7 \rho_8 \rho_9} \, , \qquad I = \text{M, N, IY, H} \;.
\label{eq:I_I}
\ee
The numerators for each topology, ${\cal N}_I$, are defined below.
The details of the calculation can be found in the ancillary files accompaning the \texttt{arXiv} submission of this article. In particular, using \texttt{xTensor} \cite{Martin-Garcia:xAct} the \texttt{Mathematica} notebook \texttt{Contractions.nb} computes  the integrand of eq.~\eqref{eq:I_I} using the stress-energy tensor and prints the results in four different text files. These files are then imported in \texttt{IBP-Basis1.nb}, which performs the needed IBP reductions using \texttt{LiteRed} \cite{Lee:2012cn, Lee:2013mka} and computes ${\cal E}_{I}(\gamma) $  for each topology.

\subsubsection{M topology}

We start from the M topology, i.e.~we solve eq.~\eqref{EI} with 
\be
{\cal N}_{\rm M}  = P_{\mu \nu\rho\sigma} t_{\firstc}^{\mu \nu}  t_{\firstc}^*{}^{\rho\sigma }   \;.
\ee
Performing the contractions and IBP reduction  with \texttt{LiteRed} \cite{Lee:2012cn, Lee:2013mka}, one can eventually write this contribution in terms of a single master integral,
\be
{\cal I}_{\text{M}} \left(\gamma, \varepsilon\right)  = C_{\text{M}} \!\left(\gamma, \varepsilon \right) g_1 \left(\gamma, \varepsilon \right)  \;,
\ee
where $C_{\text{M}}$ is a (not very illuminating) function of $\epsilon$ and $\gamma$.
Here we are interested in the limit  $\varepsilon \to 0$. Since  $C_{\text{M}}$ starts at $\varepsilon^0$, we just need $g_1$ at order $\varepsilon^0$, see eq.~\eqref{eq:sol0}. After performing the Fourier transform in $q$ in eq.~\eqref{EI} using
\be
\int_q \dd (q\cdot u_1) \dd (q\cdot u_2) e^{i q \cdot b} \sqrt{-q^2} = \frac{1}{\sqrt{\gamma^2-1}} \int \frac{d^2 \vq_{\perp}}{(2\pi)^2} e^{-i \vq \cdot \vb} |\vq_{\perp}| =  - \frac{1}{2 \pi } \frac{1}{ \sqrt{\gamma^2-1} |\vb|^3} \;, 
\ee
we obtain
\be
\label{eq:AM}
 {\cal E}_{\text{M}} (\gamma) = - \frac{\pi  }{8  } \left(\frac{20 \gamma ^7+16 \gamma ^6+12 \gamma ^4-13 \gamma ^3-24 \gamma ^2+15 \gamma +18}{3 \sqrt{\gamma ^2-1}}\right) \, .
\ee
The final result is unaltered by the exchange $1 \leftrightarrow 2$, therefore the symmetric contribution gives exactly the same result.

Note that the M topology does not contain contributions from the graviton cubic vertex and the involved Fourier-space waveform (the amplitude) can be computed exactly. In this case we can also compute the above contribution in a more ``direct'' way from eq. \eqref{Pmu}, by taking the relevant part of the amplitude from Sec.~\ref{sec:ampl}. 
Specifically,   ${\cal E}_{\text{M}}$ can be computed   from eq.~\eqref{Pmu} as 
\be 
\label{eq:AM_an}
{\cal E}_{\text{M}} = 256 \pi^3 |\vb|^3 \sum_\lambda \int_k \ddp(k^2) k \cdot (u_1 + u_2) \, {\cal A}_\lambda^{\firstc} (k) {\cal A}_\lambda^{\firstc} (- k) \, ,
\ee
with ${\cal A}_\lambda^{\firstc} (k)$ defined in \eqref{eq:Al1_MI}. Working in 4 dimensions and solving explicitly the integral in $q$ of eq. \eqref{eq:Al1_MI} as we did in App.~\ref{app:coeff}, we find the following  expression in terms of modified Bessel functions of the second kind $K_n$,
\begin{align}
{\cal A}_\lambda^{\firstc} (k) & = \frac{\epsilon^{\lambda *}_{\mu\nu}}{4\pi \sqrt{\gamma^2 -1 }}\bigg\{ \frac{u_1^{\mu} u_1^{\nu}}{\gamma^2 -1} \bigg[\left( 1 + \gamma (3 - 2\gamma^2)\frac{z_2}{z_1} \right) K_0 \left(z_1\right) - i (1-2\gamma^2) \frac{k \cdot b}{z_1} K_1 \left(z_1\right) \bigg] \notag \\
& \qquad+ 2 i \frac{u_1^\mu b^\nu}{|\vb| \sqrt{\gamma^2 -1}}\left(1-2\gamma^2\right) K_1 \left(z_1\right) - 2\frac{u_1^\mu u_2^\nu}{\gamma^2-1}\gamma(3-2\gamma^2) K_0 \left(z_1\right) \bigg\}  \, ,
\end{align}
where we have defined, for $a=1, 2$,
\begin{align}
z_a \equiv \frac{|\vb| k \cdot u_a}{\sqrt{\gamma^2-1}} \, .
\end{align}
Using again eq. \eqref{eq:sum_pol}, we can rewrite eq. \eqref{eq:AM_an} as follows
\begin{align}
{\cal E}_{\text{M}} = & \frac{|\vb|^2}{2 \pi ^2 \left(\gamma ^2-1\right)^{5/2}}\int_{0}^\infty \! d\omega \int\! d\Omega \, \omega\,\frac{ z_1+z_2}{z_1^2}\Big\{ - (1-2\gamma^2)^2 \big[\vk \cdot \vb +4(\gamma^2-1) z_1^2\big] K^2_1\big(z_1\big)\notag \\
& \quad + \big[\big(z_1 + \gamma (3-2\gamma^2) z_2 \big)^2 - 4 z_1 \gamma^2 (3-2\gamma^2) \big( z_1 - (3-2\gamma^2)  (z_1-\gamma z_2) \big) \big] K^2_0\big(z_1\big) \Big\} \, ,
\label{eq:AM_an_f}
\end{align}
where we used that
\be
{d^4 k} \, \ddp(k^2) = 2 \pi   {d^3 \vk}/(2 |\vk|)  |_{k^0 = |\vk|} = \pi  \, d\Omega \, {d\omega } \, \omega |_{k^0 = \omega} \, , \qquad \omega \equiv |\vk| \, .
\ee
Working again in the frame \eqref{eq:frame}, one can first solve the integrals in the azimuthal angle $\phi$ and the frequency $\omega$, then finally in the polar angle $\theta$, eventually recovering eq.~\eqref{eq:AM}.
We stress again that such a direct procedure is unavailable when the amplitude involves ${\cal A}_\lambda^{\vdash} (k)$ (see eq.~\eqref{eq:Ac_MI}). 

\subsubsection{N topology}

For the N topology the numerator in eq.~\eqref{EI} is
\be
{\cal N}_{\rm N}   = P_{\mu \nu\rho\sigma} t_{\firstc}^{\mu \nu}   t_{\secondc}^*{}^{\rho\sigma }  \;.
\ee
Performing again the IBP reduction procedure and using the symmetry $u_1 \leftrightarrow u_2$, we find that ${\cal I}_{\rm N}$ can be rewritten in terms of two master integrals, $g_2$ and $g_3$, 
\be 
 {\cal I}_{\rm N} (\gamma, \varepsilon )=   C_{\text{N},2}(\gamma, \varepsilon) g_2(\varepsilon,\gamma) + \frac{C_{\text{N},3}(\gamma,  \varepsilon )}{\varepsilon \sqrt{\gamma^2 - 1}} g_3 (\gamma, \varepsilon) \, ,
\ee
where $C_{\text{N},2}$ and $C_{\text{N},3}$ are functions starting at order $\varepsilon^0$.
The coefficient in front of $g_3$  diverges  for $\varepsilon \rightarrow 0$ but this is compensated by   $g_3$ that starts at order $\varepsilon$, see eqs.~\eqref{eq:sol0} and \eqref{eq:sol3_1}.   Inserting the leading order solutions for $g_2$ and $g_3$, we eventually find 
 \begin{align}
 \label{eq:AN}
 {\cal E}_{\text{N}} =  \frac{\pi  }{8  }  \bigg[ &  \frac{4 \left(20 \gamma ^6-64 \gamma ^5+98 \gamma ^4-80 \gamma ^3+28 \gamma ^2-1\right)}{\left(\gamma ^2-1\right)^{3/2}}  \notag \\
 & \quad + \frac{8 \left(4 \gamma ^6-10 \gamma ^4 + 8 \gamma ^2-3 \right)}{\left(\gamma ^2-1\right)^{3/2}} \frac{\gamma \arcsinh\left(\sqrt{\frac{\gamma - 1}{2}}\right) }{\sqrt{\gamma^2 -1 }}\bigg] \, ,
 \end{align}
 where we have used eq.~\eqref{eq:defx} to replace
 \be
 \label{eq:Arc/Log}
 -\log\left(x\right) = 2 \arcsinh\left(\sqrt{\frac{\gamma - 1}{2}}\right)  \;.
 \ee

\subsubsection{IY topology}

For the IY topology the numerator in eq.~\eqref{EI} is
\be
{\cal N}_{\rm IY}  = 2 P_{\mu \nu \rho\sigma}  \text{Re}\left[ t_{\firstc}^{\mu \nu} t_{\vdash}^*{}^{\rho\sigma }   \right]   \;.
\ee
After the IBP reduction  we find that ${\cal I}_{\rm IY}$ is given in terms of $g_1$, $g_2$ and $g_3$, i.e.
\be 
{\cal I}_{\text{IY}} =   C_{\text{IY},1} \, g_1 +C_{\text{IY},2} \, g_2 + \frac{C_{IX,3} }{\varepsilon \sqrt{\gamma^2 - 1}} \, g_3  \, ,
\ee
where the dependence on $\varepsilon$ and $\gamma$ of the functions above is understood.
Both $ C_{\text{IY},1}$ and $ C_{\text{IY},2}$ start at order $\varepsilon^{-1}$, leading to a 
 seemingly divergent term for $\varepsilon \to 0$, 
\be
{\cal I}_{\text{IY}}  \supset \frac{1}{\varepsilon}\left[\frac{2 \gamma ^4-3 \gamma ^2+3}{8 } (g_1 - g_2) -\frac{\gamma  \left(6 \gamma ^4+\gamma ^2-15\right) }{32 \sqrt{\gamma ^2-1}} g_3 \right] \, .
\ee
However, this is finite because both $g_1-g_2$ and $g_3$ start at  order  $\varepsilon$. 

Inserting the solutions for $g_1$, $g_2$ and $g_3$ given in eqs.~\eqref{eq:sol0}--\eqref{eq:sol3_1}, we eventually obtain
\begin{align}
\label{eq:AIY}
 {\cal E}_{\text{IY}} = \frac{\pi  }{8  } & \bigg[  \frac{208 \gamma ^9 \! +\!176 \gamma ^8 \! - \! 448 \gamma ^7 \! - \! 214 \gamma ^6 \! +\! 436 \gamma ^5 \! -\! 6025 \gamma ^4 \! +\! 14216 \gamma ^3 \! -\! 11108 \gamma ^2\! +\! 2676 \gamma \! +\! 83}{12 \left(\gamma ^2-1\right)^{3/2}}\notag \\
 &\!\! +\frac{ \left(6 \gamma ^4+\gamma ^2-15\right) }{2 \sqrt{\gamma ^2-1}} \frac{\gamma \arcsinh\left(\sqrt{\frac{\gamma- 1}{2}}\right) }{\sqrt{\gamma^2 -1 }} - \frac{4 \left(2 \gamma ^4-3 \gamma ^2+3\right)}{\sqrt{\gamma ^2-1}} \log\left(\frac{\gamma + 1}{2}\right)\bigg] \, ,
 \end{align}
 where we used again eq.~\eqref{eq:Arc/Log} and 
 \be 
 \log\left(\frac{(x+1)^2}{4 x}\right) = \log\left(\frac{\gamma + 1}{2}\right) \, .
 \ee

\subsubsection{H topology}

Finally, we need to compute the contribution of the H topology, for which
\be
{\cal N}_{\rm H}   \equiv\frac{1}{2} P_{\mu \nu\rho\sigma} t_{\vdash}^{\mu \nu}     t_{\vdash}^*{}^{\rho\sigma } \;.
\ee
IBP reducing one last time, we find 
\be 
{\cal I}_{\text{H}} =   C_{\text{H},1}\,  g_1 +C_{\text{H},2} \, g_2 + C_{H,4} \, g_4 \, .
\ee
Once again, the cancellation of divergencies for $\varepsilon \rightarrow 0$ is non-trivial. Before expanding $g_1$, $g_2$ and $g_4$ we obtain a seemingly divergent term, 
\begin{align}
{\cal I}_{\text{H}} \supset \frac{1}{4 \varepsilon}\bigg[ & \frac{83 \gamma ^4-420 \gamma ^3+738 \gamma ^2-532 \gamma +195}{12} g_2 -\frac{35 \gamma ^4-60 \gamma ^3+90 \gamma ^2-76 \gamma +27 }{2} g_4 \notag \\
 & \qquad \quad- \left(2 \gamma ^4-15 \gamma ^3+27 \gamma ^2-19 \gamma +7\right) g_1 \bigg] \, ,
\end{align}
which however is finite once we use the solutions for $g_1$, $g_2$ and $g_4$, eqs.~\eqref{eq:sol0}--\eqref{eq:sol2_1} and \eqref{eq:sol4_1}.
Using these, we obtain
\begin{align}
\label{eq:AH}
{\cal E}_{\text{H}} =  - \frac{\pi  }{8  }  \bigg[ &  \frac{64 \gamma ^9 \! +\! 56 \gamma ^8 \! -\! 184 \gamma ^7 \! +\! 276 \gamma ^6 \! -\!  1016 \gamma ^5 \! -\! 758 \gamma ^4 \! +\! 5588 \gamma ^3 \! -\! 6540 \gamma ^2 \! +\! 3036 \gamma \! -\! 522}{6 \left(\gamma ^2-1\right)^{3/2}}\notag \\
 & +\frac{19 \gamma ^4 + 60 \gamma ^3 - 126 \gamma ^2 + 76 \gamma - 29}{2 \sqrt{\gamma ^2-1}} \log\left(\frac{\gamma + 1}{2}\right)\bigg] \, .
 \end{align}
 
\subsection{Full result}

Summing up all the  above contributions, i.e.~eqs.~\eqref{eq:AM}, \eqref{eq:AN},\eqref{eq:AIY} and \eqref{eq:AH}, and taking into account also the symmetric ones, we eventually obtain
\be 
{\cal E}(\gamma) =\frac{\pi}{8} \left[  f_1(\gamma)+f_2(\gamma)\log\left(\frac{\gamma + 1}{2}\right) + f_3(\gamma) \frac{\gamma\, \arcsinh\left(\sqrt{\frac{\gamma -1}{2}}\right) }{\sqrt{\gamma^2 -1 }}  \right]\, ,
\ee
with
\begin{align}
f_1(\gamma)& = \frac{210 \gamma ^6-552 \gamma ^5+339 \gamma ^4-912 \gamma ^3+3148 \gamma ^2-3336 \gamma +1151}{6 \left(\gamma ^2-1\right)^{3/2}}\, ,\\
f_2(\gamma)& = -\frac{35 \gamma ^4 + 60 \gamma ^3 - 150 \gamma ^2 + 76 \gamma - 5}{\sqrt{\gamma ^2-1}}\, ,\\
 f_3(\gamma) & = \frac{  \left(2 \gamma ^2-3\right) \left(35 \gamma ^4-30 \gamma ^2+11\right)}{\left(\gamma ^2-1\right)^{3/2}}\, .
\end{align}
This result agrees with the one recently derived via other methods \cite{Herrmann:2021lqe,Herrmann:2021tct,DiVecchia:2021bdo}, and complete the leading-order radiated sector derived with an EFT worldline approch \cite{Mougiakakos:2021ckm}.

From eq.~\eqref{totalmom} one can compute the center-of-mass radiated energy, $E^{\rm (CoM)}_{\rm rad} \equiv P_{\rm rad} \cdot u_{\rm CoM}$, obtaining \cite{Herrmann:2021lqe}
\be
\label{Eradcom}
E^{\rm (CoM)}_{\rm rad} = \frac{G^3 m^4 \nu^2 }{|\vb|^3 h(\nu, \gamma)} {\cal E}(\gamma) + {\cal O}(G^4) \;,
\ee
where we have defined the total mass $m \equiv m_1+m_2$, the symmetric mass ratio $\nu \equiv m_1 m_2/m^2$ and $h (\nu,\gamma) \equiv \sqrt{1+2 \nu (\gamma-1)}$.
This result  has been used to check with the literature in different regimes. For instance, one can compare against post-Newtonian computations up to 2PN \cite{Kovacs:1978eu,Blanchet:1989cu,Bini:2020hmy} by expanding it for small velocities. 
From eq.~\eqref{Eradcom}, one can also obtain the radiated energy in elliptic orbits in the high ellipticity limit via analytic continuation \cite{Kalin:2019rwq,Kalin:2019inp} and compare the  small velocity expansion with known 3PN results \cite{Blanchet:2013haa}.
Finally, the same radiated energy also appears as a tail effect in the 4PM Hamiltonian computed in \cite{Bern:2021dqo}. We refer  to \cite{Herrmann:2021lqe,Herrmann:2021tct,Dlapa:2021npj} for a more thorough discussion.

\section{Conclusion}
\label{Sec5}

We have used the worldline approach to directly derive, for the first time using  purely classical methods,  
the four-momentum radiated during the encounter of two spinless bodies at leading-order in the post-Minkowskian expansion, i.e.~at ${\cal O}(G^3)$.  The calculation can be roughly  split into two parts. The first (Sec.~\ref{Sec2}) involves the derivation of the classical amplitude of one graviton emission, which was the subject of our previous work with Mougiakakos \cite{Mougiakakos:2021ckm}. Here we have reformulated the steps leading to the stress-energy tensor and the on-shell radiation amplitude, providing details omitted in that reference (see e.g.~Apps.~\ref{AppA} and \ref{AppB}).  In the worldline approach the two bodies are treated as non-propagating external classical sources, as far as gravitons are concerned.  In particular, quantum contributions are represented by graviton loops that we neglect. The derived stress-energy tensor and the radiation amplitude are  classical objects from the onset. In contrast with scattering-amplitude based methods, there is no need to take an exponential expansion that leads to the treatment of unphysical super-classical terms \cite{Kosower:2018adc} in the derivation.

The second part of the calculation, reported in Secs.~\ref{Sec3} and \ref{Sec4}, concerns the computation of the radiated observable. This involves a phase-space integration of a single graviton four-momentum weighed by the probability density of emission, i.e.~the modulo squared of the on-shell amplitude. Using reverse unitarity, we  treated the phase-space delta function as a cut propagator, allowing us to reformulate the integration as a two-loop Feynman integral that we  attacked using  the paraphernalia  developed in particle physics: integration-by-parts identities and differential equations. We split the integration into four topologies that naturally follow from the worldline approach, each of which can be  reduced by integration by parts to a set of only four master integrals. 
To compute these master integrals we solved the system of canonical differential equations that they satisfy.  Their boundary conditions, given in the near-static limit ($\gamma \to 1$), are  discussed in App.~\ref{AppC}.
The explicit reduction to master integrals is instead reported in the ancillary files accompanying the \texttt{arXiv} submission of this work. In would be interesting to compare the intermediate steps of our derivation with those of other amplitude- or eikonal-based methods such as those of Ref.~\cite{Kosower:2018adc,Herrmann:2021lqe,DiVecchia:2021bdo}.

An obvious future direction is the extension of the present computation to higher orders. This will require to include the effect of the interplay between radiation and potential gravitons. We also expect 
the Feynman rules and the number of diagrams to increase rapidly. The classical double copy \cite{Goldberger:2016iau,Goldberger:2017vcg,Shen:2018ebu,Shi:2021qsb} could be a promising approach to reduce these complications. Moreover, it would be useful to apply techniques analogous to those discussed in \cite{Herrmann:2021tct,Grozin:2015kna} to simplify the computation of the boundary conditions of the master integrals.
Other natural extensions, of course, include tidal/dissipative \cite{Goldberger:2020wbx,Goldberger:2020fot} and spin effects \cite{Jakobsen:2021lvp,Jakobsen:2021zvh} in the radiated four-momentum. 

Our approach is promising  also to derive  other radiated observables, such as the angular momentum at orders higher than $G^2$. More generally, radiation-reaction effects will be required to incorporate the dissipative dynamics in semi-analytic models such as the  EOB (see e.g.~\cite{Bini:2012ji,Khalil:2021txt,Bini:2021gat,Damgaard:2021rnk,Saketh:2021sri}) and will be crucial to develop accurate templates to make the most of future gravitational wave astronomy.

\subsection*{Acknowledgements}

We would like to thank Brando Bellazzini, Carlo Heissenberg, Gregor K\"{a}lin, Stavros Mougiakakos, Julio Parra-Martinez, Rafael Porto and Leong Khim Wong for insightful discussions. This work was partially supported by the CNES.

\appendix 


\section{Stress-energy tensor}\label{App:Tmunu}
\label{AppA}

Here we report the expression of the stress-energy tensor conserved off-shell. For generality, we have computed it  in $d$ spacetime dimensions. Since the cubic vertex in de Donder gauge is the same  in any dimension, the only modifications of the Feynman rules are  the graviton propagator \eqref{eq:propagator}, which becomes 
\be
\Gprop = \frac{i}{k^2}\left(\eta_{\mu(\rho}\eta_{\sigma)\nu} - \frac{1}{d-2}\eta_{\mu\nu}\eta_{\rho\sigma}\right) \, ,
\ee
and the first-order deviation of the equation of motion \eqref{deltau} and \eqref{deltax}, which are now
\begin{align}
\delta^{(1)} u_1^\mu (\tau)= & \    \frac{m_2}{4 \mpl^{d-2}}\int_q \dd(q \cdot u_2) \frac{e^{-i q\cdot b - i q\cdot u_1 \tau}}{q^2} \left[\beta \frac{q^\mu}{q\cdot u_1 +i0^+} - 2 \gamma u_2^\mu + \frac{2}{d-2}u_1^\mu \right] \;,  \\
\delta^{(1)} x_1^\mu (\tau)= & \ \frac{i m_2}{4 \mpl^{d-2}}\int_q \dd(q \cdot u_2) \frac{e^{-i q\cdot b - i q\cdot u_1 \tau}}{q^2 (q\cdot u_1 + i0^+)} \left[\beta \frac{q^\mu}{q\cdot u_1 +i0^+} - 2 \gamma u_2^\mu + \frac{2}{d-2}u_1^\mu \right] \;,
\end{align}
where
\be
\beta \equiv \gamma^2 - \frac{1}{d-2}\, .
\ee

For convenience, we define
\be
\mu_{1,2} (k) \equiv e^{i (q_1 \cdot b_1 + q_2 \cdot b_2)} \dd^{(d)}(k - q_1 -q_2) \dd(q_1 \cdot u_1) \dd(q_2 \cdot u_2) \;.
\ee
The stress-energy tensor at next-to-leading order in $G$ is given by the sum of the following terms\footnote{We thank Gregor K\"alin and Rafael Porto for several consistency checks of this result.}
\begin{align}
\label{eq:t1off}
\tilde T_{\firstc}^{\mu\nu} (k) =& \frac{m_1 m_2}{4 \mpl^{d-2}} \int_{q_1, q_2} \mu_{1,2} (k) \frac{1}{q_2^2} \bigg[ \frac{2  \beta }{k \cdot u_1} q_2^{(\mu} u_1^{\nu)} - 4 \gamma u_1^{(\mu} u_2^{\nu)} \notag \\
&\hspace{5cm}- \left[ \beta \frac{k \cdot q_2}{(k \cdot u_1)^2} - 2\gamma \frac{k \cdot u_2}{ k \cdot u_1} - \frac{2}{d - 2} \right] u_1^{\mu} u_1^{\nu} \bigg]  \, , \\
\label{eq:t2off}
\tilde T_{\secondc}^{\mu\nu} (k) =& \frac{m_1 m_2}{4 \mpl^{d-2}} \int_{q_1, q_2} \mu_{1,2} (k) \frac{1}{q_1^2} \bigg[ \frac{2  \beta }{k \cdot u_2} q_1^{(\mu} u_2^{\nu)} - 4 \gamma u_2^{(\mu} u_1^{\nu)}\notag \\
&\hspace{5cm}  - \left[ \beta \frac{k \cdot q_1}{(k \cdot u_2)^2} - 2\gamma \frac{k \cdot u_1}{ k \cdot u_2} - \frac{2}{d - 2} \right] u_2^{\mu} u_2^{\nu} \bigg]  \, , \\
\label{eq:tcoff}
\tilde T_{\vdash}^{\mu\nu} (k) =&  \frac{m_1 m_2}{4 \mpl^{d-2}} \int_{q_1, q_2} \mu_{1,2} (k)\frac{1}{q_1^2 q_2^2}\bigg\{ \beta \left( q_1^\mu q_1^\nu + q_2^\mu q_2^\nu + k^\mu k^\nu \right)  \!+\! 2\bigg(( k \cdot u_2)^2 - \frac{k^2 + q_1^2 }{d-2} \bigg) u_1^\mu u_1^\nu   \notag \\
 & \qquad + 2 \bigg( ( k \cdot u_1)^2 - \frac{k^2 + q_2^2}{d-2} \bigg) u_2^\mu u_2^\nu  + 4 \gamma ( k \cdot u_2) q_2^{(\mu}u_1^{\nu)}+ 4 \gamma ( k \cdot u_1) q_1^{(\mu}u_2^{\nu)} \notag \\
 & \qquad - \eta^{\mu\nu} \left[ 2 \frac{(k \cdot u_1)^2 + (k \cdot u_2)^2}{d - 2} -2 \gamma (k\cdot u_1) (k \cdot u_2) +\frac{\beta}{2} \left(3k^2 + q_1^2 +q_2^2 \right)\right] \notag \\
 & \qquad + 2 \left[ \gamma \left(q_1^2 +q_2^2 + k^2 \right) - 2 (k \cdot u_1) (k \cdot u_2)  \right] u_1^{(\mu} u_2^{\nu)} \notag \\
 & \qquad +4\left( \frac{k\cdot u_1}{d-2} - \gamma k \cdot u_2\right) k^{(\mu}u_1^{\nu)} + 4\left( \frac{k\cdot u_2}{d-2} - \gamma k \cdot u_1\right) k^{(\mu}u_2^{\nu)} \bigg\} \, .
 \end{align}
It can be verified that it is transverse for on-shell and off-shell gravitons as well, i.e.,
\begin{equation}
k_\mu \left(  \tilde T_{\firstc}^{\mu\nu} (k) + \tilde T_{\secondc}^{\mu\nu} (k) + \tilde T_{\vdash}^{\mu\nu} (k)\right)  = 0 \, , \quad \forall k \, .
\end{equation}
By using  momentum conservation, on-shell and harmonic gauge conditions, we recover the expressions in eqs.~\eqref{eq:t1}--\eqref{eq:tc}.

\section{Integrals involved in the amplitude}\label{app:coeff}
\label{AppB}

This appendix is devoted to solve the set of integrals defined in eqs.~\eqref{eq:MI} and \eqref{eq:MJ}. In particular, we show how  $I_{(n)}$  can be solved exactly, while  $J_{(n)}$   can be at best rewritten as an integral over a one-dimensional Feynman parameter.

Let us start with the scalar integral $I_{(0)}$. Since the final result will be in terms of Lorentz invariants, we can solve this integral in a particular frame and it is convenient to pick the one defined in \eqref{eq:frame}. Solving the two delta functions, we reduce $I_{(0)}$ to a two-dimensional integral over the components $\vq_{\perp}$ that lie on the plane perpendicular to the direction of the scattering bodies. Hence
\be
\begin{split}
I_{(0)} & = -\frac{1}{\sqrt{\gamma^2 -1}}\int \frac{d^2 \vq_\perp}{(2\pi)^2}\frac{e^{i \vq_\perp \cdot \vb}}{\vq_\perp^2 + \frac{(k \cdot u_1)^2}{\gamma^2 -1}}  \\
& = -\frac{1}{\sqrt{\gamma^2 -1}}\int_0^\infty\!\! dt \int \frac{d^2 \vq_\perp}{(2\pi)^2}\exp\left[- t \vq_\perp^2 + i \vq_\perp \cdot \vb - t\frac{(k \cdot u_1)^2}{\gamma^2 -1}\right] \, ,
\end{split}
\ee
where in the second step we introduced a Schwinger parameter $t$. Solving the Gaussian integral in $\vq_\perp$ eventually gives  the final result, i.e.,
\begin{align}
I_{(0)} & = - \frac{1}{4\pi \sqrt{\gamma^2 -1}}\int_0^\infty\!\! dt \, \frac{1}{t} \exp\left[ -\frac{|\vb|^2}{4 t} - t \frac{(k \cdot u_1)^2}{\gamma^2 -1} \right] = - \frac{K_0 \left( z_1\right)}{2\pi \sqrt{\gamma^2 -1}} \, ,
\end{align}
where $K_{n}$ are  modified Bessel functions of the second kind and we defined 
\begin{equation}
z_a \equiv \frac{\sqrt{-b^2} (k \cdot u_a)}{\sqrt{\gamma^2 -1}} \, \qquad a=1, 2 \, .
\end{equation}

Once  the scalar integral is solved, the vectorial $I_{(1)}^\mu$ can be computed decomposing it on a complete basis, i.e.
\begin{equation}
\label{eq:I1_basis}
I_{(1)}^\mu = A_b b^\mu + A_u (u_1^\mu -\gamma u_2^\mu) \, ,
\end{equation}
where the dependence on the combination $u_1^\mu -\gamma u_2^\mu$ comes from the fact that $u_2\cdot I_{(1)} = 0$. Contracting both sides with $b^\mu$ and $u_1^\mu$, one eventually obtains
\be
\begin{split}
A_b & = \frac{i b^\mu}{b^2}\frac{\partial I_{(0)}}{b^2} = -\frac{i}{2\pi \sqrt{\gamma^2 -1}}\frac{z_1 \, K_1(z_1)}{|b^2|}  \; , \\
A_u & = - \frac{k \cdot u_1}{\gamma^2 -1}I_{(0)} = \frac{k \cdot u_1}{2\pi (\gamma^2 -1)^{3/2}} K_0(z_1) \, .
\end{split}
\ee
 
The second set of integrals, defined in eq.~\eqref{eq:MJ}, is  more involved due to the presence of a second massless propagator. Starting again with the scalar integral $J_{(0)}$, we can use Feynman parametrization to rewrite it in terms of only one massless propagator,
\be
\begin{split}
J_{(0)} & = \int_0^1\!dy \int_q\dd(q\cdot u_1 - k\cdot u_1)\dd(q\cdot u_2)\frac{e^{- i q \cdot b}}{(q- y k)^4} \\
& = \int_0^1\!dy e^{- i y  k \cdot b} \int_q\dd(q\cdot u_1 - (1-y) k\cdot u_1)\dd(q\cdot u_2 + y k \cdot u_2)\frac{e^{- i q \cdot b}}{q^4} \, ,
\end{split}
\ee
where for the second line we have performed the shift $q \to q + y k$ and we have imposed the $k^2=0$, because the amplitude must be eventually  evaluated on-shell. At this point we can follow a procedure analogous to the one we used for $I_{(0)}$. Choosing again the frame  \eqref{eq:frame}, we use the two delta functions to reduce the computation to a two-dimensional integral over $\vq_\perp$, that we can solve using Schwinger parametrization. This yields
\begin{align}
\label{eq:J0_step}
J_{(0)} = \frac{1}{\sqrt{\gamma^2 -1}}\int_0^1\!dy e^{- i y  k \cdot b}\int_0^\infty\!\! dt \, t \int\frac{d^2\vq_\perp}{(2\pi)^2} \exp\left[- t \vq_\perp^2 + i \vq_\perp \cdot b - t \frac{s^2(y)}{\gamma^2-1}\right] \, ,
\end{align}
where we have defined 
\begin{equation}
s(y) \equiv \sqrt{(1-y)^2 (k\cdot u_1)^2 + 2\gamma y(1-y) (k\cdot u_1)(k\cdot u_2) +y^2 (k\cdot u_2)^2 } \, .
\end{equation}
Notice that $s(y)$ changes when computing the symmetric contribution $1 \leftrightarrow 2$. 
At this point, the integral over $\vq_\perp$ in eq.~\eqref{eq:J0_step} is Gaussian and can be easily solved,
\begin{align}
\label{eq:J0_final}
J_{(0)}  = \frac{\sqrt{-b^2}}{4 \pi }\int_0^1\!dy e^{- i y  k \cdot b}\frac{K_1\left( w(y) \right)}{s(y)} \, ,
\end{align}
where we introduced the shorthand notation
\be
w(y) \equiv \frac{\sqrt{-b^2} s(y) }{\sqrt{\gamma^2 -1}} \, . 
\ee

We can solve $J_{(1)}^\mu$ and $J_{(2)}^{\mu\nu}$  analogously  to what we did for $I_{(1)}^\mu$ with the  difference  that, before decomposing on a complete basis as in eq.~\eqref{eq:I1_basis}, we find it convenient to use again Feynman parametrization. For instance, for $J_{(1)}^\mu$ we have
\begin{align}
J_{(1)}^\mu & = \int_0^1\!dy  e^{- i y  k \cdot b} \int_q\dd(q\cdot u_1 - (1-y) k\cdot u_1)\dd(q\cdot u_2 + y k \cdot u_2)\frac{e^{- i q \cdot b}}{q^4}(q^\mu + y k^\mu) \, , 
\end{align}
where we performed again the shift $q \to q + y k $. The contribution proportional to $k^\mu$ can be computed using the result of $J_{(0)}$, while the one proportional to $q^\mu$ must be decomposed on a complete basis. The very same procedure can be carried out for $J_{(2)}^{\mu\nu}$, yielding     
\begin{align}
\label{eq:J1_c}
J_{(1)}^\mu & = \int_0^1\!dy  e^{- i y  k \cdot b} \left[ B_b b^\mu + B_1 u_1^\mu + B_2 u_2^\mu  \right] \, , \\
\label{eq:J2_c}
J_{(2)}^{\mu\nu} & = \int_0^1\!dy  e^{- i y  k \cdot b} \left[ C_{\eta}\eta^{\mu\nu} \!+\! C_b b^{\mu}b^{\nu}\! +\! C_1 b^{(\mu} u_1^{\nu)} \!+\! C_2 b^{(\mu} u_2^{\nu)} \!+\! C_3 u_1^{(\mu}u_2^{\nu)} \!+\! C_4 u_1^\mu u_1^\nu + C_5 u_2^\mu u_2^\nu  \right] \, ,
\end{align}
where we omitted (irrelenvant) terms proportional to $k^\mu$.

To find the coefficients $B_i$ in eq.~\eqref{eq:J1_c} we must solve the  system
\be 
B_b   = \frac{i b^\mu}{b^2}\frac{\partial J_{(0)}}{\partial b^\mu} \, , \quad
B_1   = \frac{(y-1) k \cdot u_1 - y \gamma k \cdot u_2}{\gamma^2 -1}  J_{(0)} \, , \quad
B_2   = \frac{y k \cdot u_2 - (y-1) \gamma k \cdot u_1}{\gamma^2 -1}  J_{(0)} \, .
\ee
Using   eq.~\eqref{eq:J0_final}, the solutions are
\be
\begin{split}
B_b & = \frac{i}{4 \pi \sqrt{\gamma^2 - 1}} K_0\left( w(y) \right) \, , \\
B_1 & = \frac{\sqrt{-b^2}}{4 \pi (\gamma^2 - 1)} \frac{ (y - 1) k \cdot u_1  - y \gamma k \cdot u_2}{s(y)} K_1\left( w(y) \right) \, , \\
B_2 & = \frac{\sqrt{-b^2}}{4 \pi (\gamma^2 - 1)} \frac{ y k \cdot u_2  - (y - 1) \gamma k \cdot u_1}{s(y)} K_1\left( w(y) \right) \, .
\end{split}
\ee

Then contracting eq.~\eqref{eq:J2_c} with  the tensor structure on the right-hand side, we obtain the following system for the $C_i$ coefficients,
\begin{align*}
& b^2 \left( C_\eta + C_b b^2 \right) = - b^\mu b^\nu \frac{\partial^2 J_{(0}}{\partial b^\mu \partial b^\nu} \; ,   \qquad  \frac{b^2}{2}\left( C_1 + \gamma C_2 \right) = - b^2 (y - 1) k \cdot u_1 B_b \; , \\
& \frac{b^2}{2}\left( \gamma C_1 +  C_2 \right) = - b^2  y k \cdot u_2 B_b \; , \quad C_\eta + \gamma C_3 + C_4 + \gamma^2 C_5 = (y - 1)^2 ( k \cdot u_1)^2 J_{(0)} \; , \\
& C_\eta + \gamma C_3 + \gamma^2 C_4 + C_5 = y^2 ( k \cdot u_2)^2 J_{(0)} \; , \addtocounter{equation}{1}\tag{\theequation}  \\
& \gamma C_\eta + \frac{C_3}{2} (\gamma^2 + 1) + \gamma ( C_4 + C_5 ) = (y - 1)  y ( k \cdot u_1)( k \cdot u_2) J_{(0)} \; , \\
& 4 C_\eta  +  b^2 C_b  +  \gamma C_3 +  C_4 + C_5 \!=\! \int_q\dd(q\cdot u_1 - (1-y) k\cdot u_1)\dd(q\cdot u_2 + y k \cdot u_2)\frac{e^{- i q \cdot b}}{q^2} \; ,
\end{align*}
where the   right-hand side of the last equation can be computed following the same procedure we used to solve $J_{(0)}$, which gives
\begin{equation}
 -\frac{1}{2 \pi \sqrt{\gamma^2 - 1} } K_0 \left( w(y) \right) \, .
\end{equation}
Solving the previous system, we finally obtain
\begin{align*}
C_\eta & = -\frac{1}{4 \pi \sqrt{\gamma^2 -1}} K_0 \left( w(y) \right) \; ,   \\
C_b & = - \frac{1}{4 \pi (\gamma^2 -1)} \frac{s(y)}{\sqrt{-b^2}} K_1 \left( w(y) \right) \; , \\
C_1 & = \frac{i}{2 \pi (\gamma^2 -1)}\frac{(y - 1) k \cdot u_1 -  y \gamma k \cdot u_2}{\sqrt{\gamma^2 -1}} K_0 \left( w(y) \right) \; , \\
C_2 & = \frac{i}{2 \pi (\gamma^2 -1)} \frac{y k \cdot u_2  -  (y - 1) \gamma k \cdot u_1}{\sqrt{\gamma^2 -1}} K_0 \left( w(y) \right) \; , \addtocounter{equation}{1}\tag{\theequation}  \\
C_3 & \! = \frac{1}{2 \pi (\gamma^2 -1)^{3/2}} \bigg\{ \gamma K_0 \left( w(y) \right) \! - \! w(y) K_1 \left( w(y) \right) \left[ \gamma \! +\! \frac{(\gamma^2 -1)}{s^2 (y)} y (y - 1) k \cdot u_1 k \cdot u_2 \right]\bigg\} \; , \\
C_4 & = \frac{1}{4 \pi (\gamma^2 -1)^{3/2}} \bigg\{ - K_0 \left( w(y) \right)   + w(y) K_1 \left( w(y) \right) \left[ 1 + \frac{(\gamma^2 -1)}{s^2 (y)} y^2 ( k \cdot u_2 )^2 \right]\bigg\} \; , \\
C_5 & \! = \frac{1}{4 \pi (\gamma^2 -1)^{3/2}} \bigg\{ \! - \! K_0 \left( w(y) \right) \! + \! w(y)  K_1 \left( w(y) \right) \left[ 1 \! + \frac{(\gamma^2 -1)}{s^2 (y)} (y - 1)^2 ( k \cdot u_1 )^2 \right]\bigg\}  \; .
\end{align*}

\section{Boundary conditions}\label{App:BC}
\label{AppC}

In this last appendix we show how to compute the master integrals defined in eqs.~\eqref{eq:g1}--\eqref{eq:g4} in the near-static limit  to obtain the boundary conditions that we wrote in eqs.~\eqref{eq:BC} and \eqref{eq:BC2}. We are going to follow closely the appendices of Refs.~\cite{Herrmann:2021tct} and \cite{DiVecchia:2021bdo}.
\begin{figure}[t]
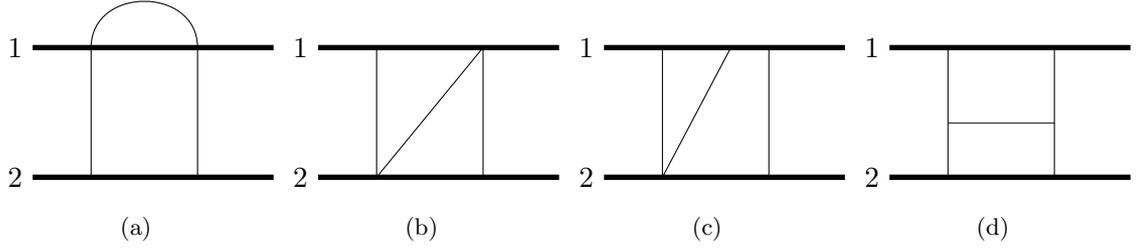

\centering
\subcaptionbox{}[.24\linewidth]{\SIone} 
\subcaptionbox{}[.24\linewidth]{\SItwo} 
\subcaptionbox{}[.24\linewidth]{\SIthree} 
\subcaptionbox{}[.24\linewidth]{\SIfour} 
\caption{Representation of the topologies of scalar integrals needed to compute the four master integrals.}
\label{fig:MI_BC}
\end{figure}

\subsection{Cutting Rules}\label{sec:cut}

Uncut master integrals are easier to solve than cut ones.
Hence,  to connect a cut master integral to its   uncut version  we can  use Cutkosky's cutting
rules \cite{Cutkosky:1960sp},\footnote{These rules are derived using Feynman and Dyson  and not retarded and advanced propagators.} as explained in App.~C of \cite{Herrmann:2021tct}. To use the cutting rules it is helpful to depict the master integrals $g_1$, $g_2$, $g_3$ and $g_4$ as diagrams. To do this, for convenience we introduce a ``propagator'' also for the massive external source, which can be seen as the soft-expanded version of the propagator of a massive scalar field \cite{Parra-Martinez:2020dzs,Herrmann:2021tct,DiVecchia:2021bdo}. Note that this is only a convenient pictorial  tool useful to solve the Feynman integral. (The compact bodies are external sources and do not propagate.) For example, for $g_1$ we have the following topology
\begin{equation}
g_1 \longrightarrow \raisebox{2pt}{\scalebox{1}{\Mcutlabel}} \, , 
\end{equation}
where a thick line denotes the ``massive propagator'' and a thin line denotes the massless one. Figure~\ref{fig:MI_BC} shows all the topologies required  to solve the four master integrals of eqs.~\eqref{eq:g1}--\eqref{eq:g4}. We then introduce the notion of scalar integrals, which are basically  Feynman diagrams in which one isolates all the factors of $i$ coming from the non-cut propagators and the factors of $-i$ coming from the vertices. To make a concrete example, let us consider again  $g_1$, represented by the diagram in Fig.~\ref{fig:MI_BC}(a),
\begin{equation}
\text{Fig.~\ref{fig:MI_BC}(a)} \ \to \ (i)^5 (-i)^4 \int_{\ell_1, \ell_2}\frac{\sqrt{-q^2}}{(2 \ell_1 \cdot u_1)^2(2 \ell_2 \cdot u_2)\ell_2^2(\ell_2 - q)^2(\ell_1+\ell_2 - q)^2} \equiv i\, { I}_1 \, .
\label{eq:SI_1}
\end{equation}
In the above equation, $ { I}_1$ is the scalar integral. 

At this point we can find a relation between $g_1$ and ${\cal I}_1$ using Cutkosky's cutting rules \cite{Cutkosky:1960sp, tHooft:1973wag, Meltzer:2020qbr}. These can be derived through  Veltman's largest time equation \cite{Veltman:1994wz} and then listed as follows:
\begin{itemize}
\item The sum of all cuts in a given channel is zero.
\item All uncut propagators and vertices on the left-hand side of the cut are unaltered, while the ones on the right-hand side are replaced by the complex conjugate of their usual expression.
\item Cut propagators are replaced by on-shell delta functions.
\end{itemize}
Given this set of rules, one can find a relation connecting cut integrals with their non-cut version. Using again the diagram depicted in Fig. \ref{fig:MI_BC}(a) as an example, we have
\begin{equation}
\label{eq:cut_bc_diag}
\Msumcut = 0 \, .
\end{equation}
In terms of scalar integrals, this relation  becomes
\begin{equation}
\label{eq:cut_bc}
(i\, {I}_1)^* + g_1 + (i\, { I}_1) = 0 \ \to \ g_1 = 2 \Imm{{ I}_1} \, .
\end{equation}
Thus,  to find the solution of the cut master integral $g_1$ in the near static limit it is enough to compute the imaginary part of the corresponding uncut scalar integral.

We can now do the same for the other three topologies depicted in Fig.~\ref{fig:MI_BC}. We have
\begin{align}
\label{eq:SI_2}
\text{Fig.~\ref{fig:MI_BC}(b)} & \ \to \ i \int_{\ell_1, \ell_2}\frac{\sqrt{-q^2}}{(2 \ell_1 \cdot u_1)^2(2 \ell_2 \cdot u_2)(\ell_1- q)^2(\ell_2 - q)^2(\ell_1+\ell_2 - q)^2} \equiv i\, {  I}_2 \, ,\\
\label{eq:SI_3}
\text{Fig.~\ref{fig:MI_BC}(c)} & \ \to \ i \int_{\ell_1, \ell_2}\frac{\sqrt{-q^2}}{(2 \ell_1 \cdot u_1)(-2 \ell_2 \cdot u_1)(2 \ell_2 \cdot u_2)(\ell_1- q)^2(\ell_2 - q)^2(\ell_1+\ell_2 - q)^2} \equiv i\, {  I}_3 \, ,\\
\label{eq:SI_4}
\text{Fig.~\ref{fig:MI_BC}(d)} & \ \to \ i\int_{\ell_1, \ell_2}\frac{(\sqrt{-q^2})^5}{(2 \ell_1 \cdot u_1)^2(2 \ell_2 \cdot u_2)\ell_1^2 \ell_2^2(\ell_1- q)^2(\ell_2 - q)^2(\ell_1+\ell_2 - q)^2} \equiv i\, {  I}_4 \, .
\end{align}
Following the same procedure as for  eqs.~\eqref{eq:cut_bc_diag} and \eqref{eq:cut_bc}, we eventually find
\begin{align}
\label{eq:cut_BC_2}
g_2 & = 2 \Imm{{ I}_2} \, , \\
\label{eq:cut_BC_3}
g_3 & = \big(\varepsilon \sqrt{\gamma^2 -1} \big) 2\Imm{{ I}_3} - \varepsilon \sqrt{\gamma^2 -1}  \VIcut \, , \\
\label{eq:cut_BC_4}
g_4 & = \frac{\gamma - 1}{8} 2 \Imm{{ I}_4}+ \frac{1- 2\varepsilon (2 +3 \gamma)}{12 (1 + 2\varepsilon)} 2 \Imm{{  I}_2} + \frac{2\varepsilon}{(1+ 2\varepsilon) (1 + \gamma)} 2\Imm{{  I}_1} \, .
\end{align} 
We can use the near-static limit of eqs.~\eqref{eq:cut_bc}, \eqref{eq:cut_BC_2}, \eqref{eq:cut_BC_3}  and \eqref{eq:cut_BC_4} to find the boundary conditions of the master integrals $g_i$. Let us focus now on solving the scalar uncut integrals $I_i$ in the near static limit.

\subsection{Integrals in the near-static limit}

In what follows, unless stated otherwise we will always consider an implicit Feynman prescription $+i 0^+$ for all the propagators.

\subsubsection*{Integral $g_1$ }

Let us start by ${ I}_1$ defined in eq.~\eqref{eq:SI_1}. Sending $\ell_1 \to \ell_1 + q - \ell_2$ we can separate the integration in $\ell_1$ and $\ell_2$
\begin{equation}
{ I}_1 = \int_{\ell_2}\frac{\sqrt{-q^2}}{(2\ell_2 \cdot u_2)\ell_2^2 (\ell_2-q)^2}\int_{\ell_1}\frac{1}{\ell_1^2 (2 \ell_1 \cdot u_1 - 2 \ell_2\cdot u_1)^2} \, .
\end{equation}
We can solve the integral in $\ell_1$ using eq. (10.25) of \cite{Smirnov:2012gma},  
\begin{equation}
{ I}_1 = -\frac{i}{(4\pi)^{2-\varepsilon}}\Gamma(1-\varepsilon)\Gamma(2\varepsilon)\int_{\ell}\frac{\sqrt{-q^2}}{(-2\ell \cdot u_1)^{2\varepsilon}(2\ell \cdot u_2)\ell^2 (\ell-q)^2} \, .
\end{equation}
For simplicity, we now perform a Wick rotation to Euclidean space, i.e., for each vector $v^\mu = (v^0, \vv) = (i v_E^0, \vv_E)$, and we use the metric $\eta_E^{\mu\nu} = \text{diag}(+, +, +, +)$ to contract the indices. The above equation becomes 
\begin{equation}
{ I}_1 = \frac{\sqrt{q_E^2}}{(4\pi)^{2-\varepsilon}}\Gamma(1-\varepsilon)\Gamma(2\varepsilon)\int_{\ell_E}\frac{1}{(2\ell_E \cdot u^E_1)^{2\varepsilon}(-2\ell_E \cdot u^E_2)\ell_E^2 (\ell_E-q_E)^2} \, .
\label{eq:I1_step}
\end{equation}
Notice that 
\begin{align}
q_E^2 = -q^2 \, , & & u_1^E \cdot u_2^E = - \gamma \, ,& & u^E_1 \cdot u_1^E = -1 = u^E_2 \cdot u_2^E \, .
\end{align}
Using Schwinger parametrization we can rewrite the integral over $\ell_E$ of eq.~\eqref{eq:I1_step} as a Gaussian integral, i.e.,
\begin{align}
{  I}_1 & = \frac{\sqrt{q_E^2}}{(4\pi)^{2-\varepsilon}}\Gamma(1-\varepsilon)\int_{\Real_+^4}\!\! d t_1 d t_2 d s_1 d s_2 \, t^{2\varepsilon -1}\notag \\
& \hspace{2.5cm} \times\int_{\ell_E}\exp\left[-t_1 (2\ell_E \cdot u^E_1) -t_2 (-2\ell_E \cdot u^E_2) -s_1\ell_E^2 -s_2 (\ell_E-q_E)^2\right] \notag \\
& = \frac{\sqrt{-q^2}}{(4\pi)^{4-2\varepsilon}}\Gamma(1-\varepsilon)\!\!\int_{\Real_+^4}\!\! d t_1 d t_2 d s_1 d s_2\frac{t_1^{2\varepsilon-1}}{(s_1+s_2)^{2-\varepsilon}}\notag \\
& \hspace{2.5cm} \times \exp\left[-\frac{s_1 s_2}{s_1+s_2}(-q^2)-\frac{t_1^2+t_2^2 -2 \gamma t_1 t_2}{s_1+s_2}\right] \, .
\end{align}
Finally, for $a=1, 2$, we can split the integrations in $t_a$ and $s_a$, by simply performing the shift $t_a \to \sqrt{s_1+s_2} t_a$, obtaining
\begin{equation}
{  I}_1 = \frac{\sqrt{-q^2}}{(4\pi)^{4-2\varepsilon}}\Gamma(1-\varepsilon)\!\!\int_{\Real_+^2}\!\!  d s_1 d s_2\frac{ e^{-\frac{s_1 s_2}{s_1+s_2}(-q^2)}}{(s_1+s_2)^{\frac{3}{2}-2\varepsilon}} \!\!\int_{\Real_+^2}d t_1 d t_2\,  t_1^{2\varepsilon-1}e^{-[t_1^2+t_2^2 -2 \gamma t_1 t_2]} \; .
\end{equation}

The integration over $s_1$ and $s_2$ can be performed using standard integration over Feynman parameters. Making the change of variables $s = s_1+ s_2$, $\tilde{s}=s_1/s$ one gets
\begin{align}
\label{eq:ints}
\int_{\Real_+^2}\!\!  d s_1 d s_2\frac{ e^{-\frac{s_1 s_2}{s_1+s_2}(-q^2)}}{(s_1+s_2)^{\frac{3}{2}-2\varepsilon}} & = \int_0^1\!\! d\tilde{s}\int_{0}^{\infty} \frac{e^{-s [\tilde{s}(1-\tilde{s})(-q^2)]}}{s^{\frac{1}{2}-2\varepsilon}} = \frac{16^\epsilon \sqrt{\pi}}{(-q^2)^{\frac{1}{2} + 2\varepsilon}}\frac{\Gamma\left(\frac{1}{2}+2\varepsilon\right)\Gamma\left(\frac{1}{2}-2\varepsilon\right)}{\Gamma(1-2\varepsilon)} \, .
\end{align}
The integration over $t_1$ and $t_2$ is a bit more delicate. Changing again variables as follows $t_2 = t\, t_1$,  one can solve the integration over $t_1$,
\begin{align}
\!\!\int_{\Real_+^2}d t_1 d t_2\,  t_1^{2\varepsilon-1}e^{-[t_1^2+t_2^2 -2 \gamma t_1 t_2]} & = \int_0^\infty \!\!d t \int_0^\infty\!\! d t_1\,  t_1^{2\varepsilon}e^{-t_1^2[1+t^2 -2 \gamma t]} \notag \\
& = \frac{\Gamma\left(\frac{1}{2} + \varepsilon\right)}{2} \int_0^\infty \!\!d t\frac{1}{(1 + t^2 - 2\gamma t)^{\frac{1}{2}+\varepsilon}} \, .
\end{align} 
Note that the integrand in $t$ is divergent for
$t = \gamma - \sqrt{\gamma^2-1}= x$ and $t = \gamma + \sqrt{\gamma^2-1}$, so one must treat it with care. In the near static limit $x \to 1$ we obtain
\begin{align}
\int_0^\infty \!\!d t\frac{1}{(1 + t^2 - 2\gamma t)^{\frac{1}{2}+\varepsilon}} & = -\frac{1}{2\varepsilon}+\frac{\sqrt{\pi}}{(1 - x)^{2\varepsilon}}\frac{\Gamma\left(\frac{1}{2}-\varepsilon\right)}{\Gamma(1-\varepsilon)}\cos(\pi \varepsilon)\big( \cot(\pi \varepsilon) - i \big) +{\cal O}({1-x}) \;.
\end{align}

Putting all together we arrive to our final result for ${ I}_1$ in the near static limit, i.e.,
\begin{align}
{  I}_1 & = \frac{1}{2 (4\pi)^{4-2\varepsilon}}\frac{\Gamma\left(\frac{1}{2}+2\varepsilon\right)\Gamma\left(\frac{1}{2}-2\varepsilon\right)}{(-q^2)^{2\varepsilon} \Gamma(1-2\varepsilon)}\bigg[ -\frac{16^\varepsilon \sqrt{\pi}}{2\varepsilon} \Gamma\left(\frac{1}{2}+\varepsilon\right)\Gamma(1-\varepsilon) \notag \\
\label{eq:I1_noncut}
&\qquad + \frac{16^\varepsilon \pi }{(1-x)^{2\varepsilon}}\Gamma\left(\frac{1}{2}+\varepsilon\right)\Gamma\left(\frac{1}{2}-\varepsilon\right)\cos(\pi \varepsilon)\big( \cot(\pi \varepsilon) - i \big)\bigg] +{\cal O}({1-x})  \, .
\end{align}
Using eq.~\eqref{eq:cut_bc} we can finally find the boundary condition for the master integral $g_1$, i.e.,
\begin{equation}
g_1\big|_{\gamma \to 1} = -\frac{C_{\text{BC}}}{(4\pi)^{4-2\varepsilon}} \, ,
\end{equation}
where $C_{BC}$ has been defined in \eqref{eq:BC2} and we used that
\begin{equation}
16^\varepsilon \pi \frac{\Gamma\left(\frac{1}{2}+\varepsilon\right)\Gamma\left(\frac{1}{2}-\varepsilon\right)\cos(\pi \varepsilon)}{\Gamma(1-2\varepsilon)} = \frac{\sqrt{\pi}}{\varepsilon}\frac{\Gamma(1+\varepsilon)\Gamma(1-\varepsilon)\Gamma\left(\frac{1}{2}-2\varepsilon\right)}{\Gamma(1 - 4\varepsilon)}\sin(\pi \varepsilon) \, .
\end{equation}

\subsubsection*{Integral $g_2$ }

Let us now analyse the second scalar integral ${  I}_2$ defined in eq.~\eqref{eq:SI_2}. First of all, we perform the shift $\ell_1 \to \ell_1 + q$ and then go again to Euclidean space for simplicity,
\begin{equation}
{  I}_2 = \sqrt{q_E^2}\int_{\ell_1^E \ell_2^E}\frac{1}{(2 \ell^E_1 \cdot u^E_1)^2( - 2 \ell^E_2 \cdot u^E_2)(\ell^E_1)^2(\ell^E_2 - q_E)^2(\ell^E_1+\ell^E_2)^2} \, .
\end{equation}
Using  Schwinger parametrization and then solving the two Gaussian integrals, one eventually arrives to 
\be
\begin{split}
{  I}_2 & = \sqrt{q_E^2}\int_{\Real_+^5}\!\!dt_1\dots dt_5\int_{\ell_1^E \ell_2^E}\, t_4 \exp\Big[-t_1 (\ell_1^E)^2 - t_2 (\ell_2^E-q_E)^2 - t_3 (\ell_1^E + \ell_2^E)^2  \\
& \qquad\qquad\qquad\qquad\qquad -t_4 ( 2\ell_1^E \cdot u^E_1) - t_5 ( - 2\ell_2^E \cdot u^E_2)\Big]   \\
& = \frac{\sqrt{-q^2}}{(4\pi)^{4-2\varepsilon}}\int_{\Real_+^5}\!\!dt_1\dots dt_5\, \frac{t_4}{T^{2-\varepsilon}}\exp\bigg[-\frac{t_1 t_2 t_3}{T}(-q^2) - \frac{t_{13} t_4^5 +t_{23} t_5^2 -2 \gamma t_3 t_4 t_5}{T}\bigg] \, ,
\end{split}
\ee
where we have defined \cite{DiVecchia:2021bdo}
\begin{align}
t_{13} \equiv t_1 + t_3 \, , \qquad t_{23} \equiv t_2 + t_3 \, , \qquad T \equiv t_1 t_2 + t_1 t_3 + t_2 t_3 \, .
\end{align}
Now we shift $t_4 \to \sqrt{T} t_4$ and $t_5 \to \sqrt{T} t_5$,  splitting the computation in two integrals
\begin{equation}
{  I}_2 = \frac{\sqrt{-q^2}}{(4\pi)^{4-2\varepsilon}}\int_{\Real+^3}\!\! dt_1 dt_2 dt_3 \frac{e^{-\frac{t_1 t_2 t_3}{T}(-q^2)}}{T^{\frac{1}{2}-\varepsilon}}\int_{\Real+^2}\!\! dt_4 dt_5\, t_4 e^{-[t_{13}t_4^2 + t_{23}t_5^2 -2 \gamma t_3 t_4 t_5]} \, .
\end{equation}
The integral in $t_4$ and $t_5$ can be solve exactly. Changing variables $t_5 = t\, t_4$ we get
\begin{align}
\int_{\Real+^2}\!\! dt_4 dt_5\, t_4 & e^{-[t_{13}t_4^2 + t_{23}t_5^2 -2 \gamma t_3 t_4 t_5]} = \int_0^\infty\!\! dt \int_0^\infty\!\!dt_5\, t_4^2 e^{-t_4^2[t_{23} t^2 + t_{13} -2 \gamma t_3 t]} \notag \\
& = \frac{\sqrt{\pi}}{4}\int_0^\infty\!\! dt \frac{1}{(t^2 t_{23} + t_{13}-2\gamma t_3 t)^{\frac{3}{2}}} = -\frac{\sqrt{\pi}}{4 \sqrt{t_{13}}}\frac{\sqrt{T+t_3^2}+\gamma t_3}{(\gamma^2 -1)t_3^2 - T} \, ,
\end{align}
so that 
\begin{equation}
\label{eq:I2_step}
{  I}_2 = -\frac{\sqrt{\pi}}{4}\frac{\sqrt{-q^2}}{(4\pi)^{4-2\varepsilon}}\int_{\Real+^3}\!\! dt_1 dt_2 dt_3 \frac{e^{-\frac{t_1 t_2 t_3}{T}(-q^2)}}{T^{\frac{1}{2}-\varepsilon}}\frac{1}{\sqrt{t_{13}}}\frac{\sqrt{T+t_3^2}+\gamma t_3}{(\gamma^2 -1)t_3^2 - T} \, .
\end{equation}

At this point we need to take the static limit $\gamma \to 1$. 
One possible way  is to assume that  the three Schwinger parameters do not scale with $\gamma$, i.e.,
\begin{align}
\label{eq:ord_sc}
t_1 \, , t_2 \, , t_3 \sim \Ord{\gamma^0} \, .
\end{align}
However, in this case the resulting solution to the integral is real and, in view of eq.~\eqref{eq:cut_BC_2}, cannot contribute to the boundary conditions of $g_2$. 
In particular,  eq.~\eqref{eq:I2_step} shows that the assumption \eqref{eq:ord_sc} does not capture the integration region coming from large values of $t_3$. 
We can choose the following scaling instead \cite{DiVecchia:2021bdo}
\begin{align}
t_1 \, t_2 \sim \Ord{\gamma^0} \, , \qquad t_3 \sim  \Ord{\gamma^2 -1} \, .
\end{align}
In this limit the integration over $t_3$ factorizes so that
\begin{equation}
{  I}_2 \simeq  -\frac{\sqrt{\pi}}{2(\gamma^2 - 1)}\frac{\sqrt{-q^2}}{(4\pi)^{4-2\varepsilon}}\int_{\Real+^2}\!\! dt_1 dt_2\, \frac{e^{-\frac{t_1 t_2}{t_{12}}(-q^2)}}{t_{12}^{\frac{1}{2}-\varepsilon}}\int_0^\infty\!\! dt_3 \frac{1}{t_3^{1-\varepsilon}\left(t_3 -\frac{t_{12}}{\gamma^2 -1}\right)} \, .
\end{equation}
Changing variables,  $t_3 = z t_{12}/(\gamma^2 -1)$, we have 
\begin{equation}
{  I}_2 \simeq  -\frac{\sqrt{\pi}}{2(1-x)^{2\varepsilon}}\frac{\sqrt{-q^2}}{(4\pi)^{4-2\varepsilon}}\int_{\Real+^2}\!\! dt_1 dt_2\, \frac{e^{-\frac{t_1 t_2}{t_{12}}(-q^2)}}{t_{12}^{\frac{3}{2}-2\varepsilon}}\int_0^\infty\!\! d z \frac{1}{z^{1-\varepsilon}\left(z -1\right)} \, ,
\end{equation}
where we used that for $\gamma \to 1$, $(\gamma^2-1)^\varepsilon\sim (1-x)^{2\varepsilon}$. The integral in $z$ can be solved exactly, again taking care of the divergences in $0$ and $1$, obtaining
\begin{equation}
\int_0^\infty\!\! d z \frac{1}{z^{1-\varepsilon}\left(z -1\right)} = (-1)^{1-\varepsilon}\frac{\Gamma(1-\varepsilon)\Gamma(1+\varepsilon)}{\varepsilon} \, .
\end{equation}
Instead, the integral in $t_1$ and $t_2$ can  be solved following a procedure completely analogous to the one of eq.~\eqref{eq:ints}, i.e.~changing variables to $t_{12} = t_1 + t_2 $ and $\tilde{t} = t_1/t_{12}$. One  eventually obtains
\begin{equation}
\int_{\Real+^2}\!\! dt_1 dt_2\, \frac{e^{-\frac{t_1 t_2}{t_{12}}(-q^2)}}{t_{12}^{\frac{3}{2}-2\varepsilon}} = \frac{\Gamma\left(\frac{1}{2}+2\varepsilon\right)\Gamma\left(\frac{1}{2}-2\varepsilon\right)^2}{(-q^2)^{\frac{1}{2}+2\varepsilon} \Gamma(1-4\varepsilon)}\, .
\end{equation}

Putting these results together and using eq.~\eqref{eq:cut_BC_2}, we finally arrive to
\begin{equation}
g_2\big|_{\gamma\to 1} =  -\frac{C_{\text{BC}}}{(4\pi)^{4-2\varepsilon}} \, .
\end{equation}

\subsubsection*{Integral $g_3$ }

As shown by eq.~\eqref{eq:cut_BC_3}, finding the boundary condition of $g_3$ requires the solution of ${\cal I}_3$ in the near static limit and also the computation of another cut of figure~\ref{fig:MI_BC}(c). First of all, following a procedure analogous to what shown for ${  I}_1$ and ${  I}_2$, we find
\begin{equation}
\label{eq:I3_b_noncut}
{  I}_3\sqrt{\gamma^2-1}\big|_{\gamma=1} = \frac{i 2^{-2+2\varepsilon}\pi^2}{(4\pi)^{4-2\varepsilon}(-q^2)^{2\varepsilon}}\frac{1}{\varepsilon}\frac{\Gamma\left(\frac{1}{2}-\varepsilon\right)\Gamma\left(\frac{1}{2}-2\varepsilon\right)\Gamma\left(\frac{1}{2}+2\varepsilon\right)}{\Gamma\left(\frac{1}{2}-3\varepsilon\right)} \, .
\end{equation}

To compute $g_3$ we have to subtract the last term on the right-hand side of eq.~\eqref{eq:cut_BC_3}.
Following the rules described in section ~\ref{sec:cut}, we find
\begin{equation}
\label{eq:I3_b_step}
\varepsilon \sqrt{\gamma^2 -1}  \VIcut = - i \varepsilon \sqrt{\gamma^2 -1} \sqrt{-q^2}\int_{\ell_2} \dd(2\ell_2\cdot u_2)\dd(2\ell_2\cdot u_1) I_L I_R \, ,
\end{equation} 
where we have defined 
\begin{align}
I_L & = \frac{1}{2}\int_{\ell_1}\frac{1}{(2\ell_1\cdot u_1)(\ell_1 - q)^2(\ell_1 + \ell_2 - q)^2} \, , \\
I_R & = -\frac{1}{(\ell_2 - q)^2} \, .
\end{align}
The integral $I_L$ is a simple one-loop computation that can be carried out straightforwardly using Schwinger or Feynman parametrization, obtaining
\begin{equation}
I_L = - \frac{i}{(4\pi)^{2-\varepsilon}}\frac{2^{-1+2\varepsilon} \pi }{(-\ell_1^2)^{\frac{1}{2}-\varepsilon}}\frac{\Gamma\left(\frac{1}{2}-\varepsilon\right)\Gamma\left(\frac{1}{2}+\varepsilon\right)}{\Gamma(1-\varepsilon)} \, .
\end{equation}
Inserting everything in eq. \eqref{eq:I3_b_step} and solving the two delta functions, one eventually arrives to
\begin{align}
\varepsilon \sqrt{\gamma^2 -1}  \VIcut & = - \frac{2^{-1+2\varepsilon}}{(4\pi)^{2-\varepsilon}}\frac{\Gamma\left(\frac{1}{2}-\varepsilon\right)\Gamma\left(\frac{1}{2}+\varepsilon\right)}{4 \Gamma(1-\varepsilon)} \int\!\!\frac{d^{2-\varepsilon} {\bf \ell}_2}{(2\pi)^{2-\varepsilon}}\frac{1}{({\bf \ell}_2)^{\frac{1}{2}+\varepsilon}({\bf \ell}_2+ {\bf q})^2} \notag \\
\label{eq:I3_b_cut2}
& =\frac{2^{-1+2\varepsilon}\pi^2}{(4\pi)^{4-2\varepsilon}(-q^2)^{2\varepsilon}}\frac{1}{\varepsilon}\frac{\Gamma\left(\frac{1}{2}-\varepsilon\right)\Gamma\left(\frac{1}{2}-2\varepsilon\right)\Gamma\left(\frac{1}{2}+2\varepsilon\right)}{\Gamma\left(\frac{1}{2}-3\varepsilon\right)} \, .
\end{align}
Using eqs.~\eqref{eq:I3_b_noncut} and \eqref{eq:I3_b_cut2} into \eqref{eq:cut_BC_3}, we get
\begin{equation}
g_3\big|_{\gamma \to 1} = 0 \, .
\end{equation}
\subsubsection*{Integral $g_4$ }

Finally, we discuss the boundary condition for $g_4$. Because of the factor $\gamma - 1$ in front of the first term of eq.~\eqref{eq:cut_BC_4}, ${  I}_4$ does not contribute to the boundary condition of $g_4$, and  we do not need to compute it. Using the results computed before for ${  I}_1$ and ${  I}_2$, one can take  the near-static limit of \eqref{eq:cut_BC_4}, obtaining
\begin{equation}
g_4\big|_{\gamma \to 1} = - \frac{1}{12} \frac{C_{\text{BC}}}{(4\pi)^{4-2\varepsilon}} \, .
\end{equation}


\bibliographystyle{JHEP}
\bibliography{ref}

\end{document}